\documentclass[a4paper]{article}
\usepackage[a4paper,top=3cm,bottom=2cm,left=2.5cm,right=2.5cm,marginparwidth=1.75cm]{geometry}
\usepackage{amsfonts}

\usepackage{amssymb}
\usepackage{color}

\usepackage[all]{xy}
\usepackage{graphicx}
\usepackage{braket}
\usepackage{amsmath}
\usepackage{appendix}
\usepackage{graphicx}
\usepackage[colorinlistoftodos]{todonotes}
\usepackage{amsthm}
\usepackage{mathrsfs}
\usepackage{amssymb}
\usepackage{accents}
\usepackage{extarrows}
\usepackage{makecell}
\usepackage{bm}
\usepackage{authblk}
\usepackage{url}
\usepackage{hyperref}
\usepackage{cite}
\usepackage{eufrak}
\hypersetup{colorlinks,
            citecolor=black,
            linkcolor=black,
            urlcolor=green,
            pdftex}

\usepackage[english]{babel}
\usepackage[utf8x]{inputenc}
\usepackage[T1]{fontenc}

\title{General geometric operators in all dimensional loop quantum gravity}
\author[1]{Gaoping Long \footnote{201731140005@mail.bnu.edu.cn}}
\author[1]{Yongge Ma \footnote{mayg@bnu.edu.cn}\thanks{corresponding author}}

\affil[1]{Department of Physics, Beijing Normal University, Beijing 100875, China}

\date{}

\begin{document}
\maketitle
\begin{abstract}
Two strategies for constructing general geometric operators in all dimensional loop quantum gravity are proposed. The different constructions are mainly come from the two different regularization methods for the de-densitized dual momentum, which play the role of building block for the spatial geometry. The first regularization method is a generalization of the regularization of the length operator in standard $(1+3)$-dimensional loop quantum gravity, while the second method is a natural extension of those for standard (D-1)-area and D-volume operators. Two versions of general geometric operators to measure arbitrary $m$-areas are constructed, and their properties are discussed and compared. They serve as valuable candidates to study the quantum geometry in arbitrary dimensions.\\

\end{abstract}

\section{Introduction}
 As a non-perturbative and background-independent approach to unify general relativity (GR) and quantum physics, loop quantum gravity (LQG) has made remarkable progress \cite{Ashtekar2012Background}\cite{Han2005FUNDAMENTAL} \cite{thiemann2007modern}\cite{rovelli2007quantum}. An important prediction of this theory is the quantum discreteness of spatial geometry at Planck scale, since the spectrums of  the geometric operators, such as volume and area, are discrete \cite{rovelli1995discreteness}\cite{ashtekar1997quantumi}\cite{ashtekar1997quantumii}. A key step in the procedure of constructing these geometric operators is to regularize the classical geometric quantities in terms of holonomy and flux which have direct quantum analogues. Different choices of regularization strategies may lead to different versions of a geometric operator, e.g., the two versions of volume operator \cite{rovelli1995discreteness}\cite{ashtekar1997quantumii}\cite{lewandowski1997volume}. Some consistency checks \cite{giesel2006consistencyi}\cite{giesel2006consistencyii}\cite{yang2019consistency} on different regularization methods have been done in order to choose suitable construction and fix the regularization ambiguity. It turns out that many geometric quantities, including length \cite{thiemann1998length}\cite{bianchi2009length}\cite{ma2010new}, area \cite{rovelli1995discreteness}\cite{ashtekar1997quantumi}, volume \cite{rovelli1995discreteness}\cite{ashtekar1997quantumii}, angle \cite{major1999operators}, metric components \cite{Ma:2000au}, and spatial Riemann curvature scalar \cite{alesci2014curvature}, have been quantized as well-defined operators in the kinematic Hilbert space of LQG \cite{ashtekar1997quantumi}\cite{brunnemann2006simplification}\cite{brunnemann2008properties}\cite{loll1996spectrum}. The starting point of LQG is the Ashtekar-Barbero connection dynamics of (1+3)-dimensional GR. However, this Hamiltonian connection formulation depends on the dimensions of space such that the internal gauge group is $SU(2)$, since its definition representation and adjoint representation have same dimension. This structure can not be directly extended to higher dimensional case. An alternative connection dynamics of GR in arbitrary (D+1)-dimensions was proposed by Bodendorfer, Thiemann and Thurn \cite{bodendorfer2013newi}\cite{bodendorfer2013newiii}. In this alternative Hamiltonian connection formulation, an extra simplicity constraint is introduced and the internal gauge group is chosen as $SO(D+1)$. In the present paper, the construction of geometric operators in arbitrary dimensional LQG based on this alternative formalism will be studied.

The phase space of the classical theory is coordinatized by a canonical pair $(A_{a}^{IJ}, \pi^{b}_{KL})$ with non-trivial Poission bracket $\{A_{a}^{IJ}(x),\pi^{b}_{KL}(y)\}=2\kappa\beta\delta_{a}^{b}\delta^{[I}_K\delta^{J]}_L\delta^{(D)}(x-y)$, where $\kappa=16\pi G^{(D+1)}$ is the gravitational constant in (1+D)-dimensional space-time, $\beta$ is the Barbero-Immirzi parameter in this theory, the spatial indices read $a,b,c,...\in\{1,...,D\}$,  internal indices read $I,J,K,...\in\{1,...,D+1\}$ and $x,y,...$ are coordinates on a D-dimensional spatial manifold $\sigma$. This phase space is subject to the Gaussian constraint and simplicity constraint which induce gauge transformations, as well as spatial diffeomorphism constraint and Hamiltonian constraint which give the spacetime diffeomorphism transformations. This classical connection theory can be quantized following the standard loop quantization methods, and the resulting all dimensional LQG is equipped with a kinematic Hilbert space $\mathcal{H}=L^2(\bar{A}, d\mu_0)$ and the corresponding quantum constraints. In comparison with the standard LQG in (1+3)-dimensions, a subtle issue in the arbitrary dimensional LQG is how to solve the simplicity constraint \cite{bodendorfer2013implementation}\cite{Gaoping2019coherentintertwiner}. Besides, the construction of geometric operators in $\mathcal{H}$ is also a complicated issue. Although the candidates of (D-1)-area operator and D-volume operator in (1+D)-dimensional LQG were proposed following the same procedure in the construction of 2-area operator and 3-volume operator in standard (1+3)-dimensional LQG, a systematic method to construct more general geometric operators is lacking. The general construction method is crucial, since there are more and more geometric quantities as the increasing of spatial dimensions. Notice that for a given geometric quantity, several different classical expressions with the basic conjugate variables could exist. Hence there would be different ways to construct the corresponding geometric operators. It should also be noted that the spatial metric $q_{ab}$ is determined by the momentum variable by
  \begin{equation}\label{qab}
  q_{ab}\hat{=}\frac{1}{2}\sqrt{q}\pi_{aIJ}\sqrt{q}\pi_b^{IJ},
  \end{equation}
   where $\hat{=}$ represents ``equal to on the  simplicity constraint surface", $q$ denotes the determinant of $q_{ab}$ and $\pi_a^{IJ}$ is the inverse of $\pi^{b}_{KL}$ satisfying $\frac{1}{2}\pi^{a}_{IJ}\pi_{b}^{IJ}=\delta^a_b$. Therefore, if one could construct a basic operator corresponding to the de-densitized dual momentum $\sqrt{q}\pi_{a}^{IJ}$, a building block for all geometric operators in arbitrary dimensions would be ready. In this paper, two strategies to construct such a building block operator will be proposed. In the first strategy, we employ the expression
    \begin{equation}\label{qpi1}
   \sqrt{q}\pi_{a}^{IJ}(x)\hat{=}\frac{(D-1)}{\beta\kappa}\{A_a^{IJ}(x),V(x,\square)\},
    \end{equation}
  where $V(x,\square):=\int_{\square}d^Dy\sqrt{q}(y)$ and $\square\ni x$ is a proper small open D-dimensional region.
   In the second strategy, $\sqrt{q}\pi_{a}^{IJ}$ is purely expressed by conjugate momentum $\pi^{a}_{IJ}$.

This paper is organized as follows. In section 2, we will construct general geometric operators by the first strategy mentioned above.  Firstly, Thiemann's construction of length operator in standard (1+3)-dimensional LQG will be extended to construct a length operator in all dimensional case. Then, we will construct an alternative 2-dimensional area operator by using co-triad as building blocks in the standard (1+3)-dimensional LQG. The method can naturally be extended to construct a 2-area operator in (1+D)-dimensional LQG. Finally, following the construction procedure of the 2-area operator, by using the de-densitized dual momentum as building blocks, general $m$-area operators for $m$-dimensional surfaces in D-dimensional space will be proposed. In section 3, certain special cases of the general ``$m$-area'' operators and the problems related to their construction will be discussed. The consistency of the alternative flux operator, which is used to construct the general $m$-area operators, with the standard flux operator will also be checked. The second strategy to construct general geometric operators will be discussed in section 4. The de-densitized dual momentum is totally given by the conjugate momentum in this strategy. By suitable regularization, its components can be expressed in terms of flux and volume properly. Then it becomes an operator by replacing the flux and volume by their quantum analogues. By using this well-defined dual momentum operator as building blocks, we will get the general geometric operators corresponding to the $m$-areas which are totally composed with flux operator and volume operator. Certain special cases of the general geometric operators and their virtues and problems will be also discussed.  Our results will be summarized and discussed in the final section.

As two frameworks of connection dynamics are involved, it is necessary to explain the variables and indices appeared in this paper. We  denote by $A_a^i$, $E^b_j$, and $e_a^i$ as the Ashtekar-Barbero connection, density triad, and co-triad respectively in (1+3)-dimensional standard LQG, where $q_{ab}=e_a^ie_{bi}$ is the spatial metric in this formulation. We denote by $A_{a}^{IJ}$, $\pi^{b}_{KL}$, and $\sqrt{q}\pi_{a}^{IJ}$ as the connection, conjugate momentum, and de-densitized dual momentum respectively in all dimensional LQG.

\section{Geometric operator in all dimensional LQG: First strategy}
In the standard (1+3)-dimensional LQG the volume operator, area operator and angle operator were directly constructed by the basic flux operator $\hat{E}^{i}(S)$, while the length operators were constructed in several different ways \cite{thiemann1998length} \cite{bianchi2009length} \cite{ma2010new}. The construction of these length operators involves two steps, the classical length of a curve is expressed by co-triad $e_a^i$ in the first step. In the second step, different expressions for $e_a^i$ are used in different ways. In Thiemann's construction of length operator, $e_a^i$ is expressed as \cite{thiemann1998length}
\begin{equation}\label{stane}
e_a^i(x)=\frac{2}{\kappa\gamma}\{A_a^i(x),V(x,\square)\},
\end{equation}
 while in the other construction it is expressed as \cite{bianchi2009length} \cite{ma2010new}
 \begin{equation}\label{stane2}
 e_a^i(x)=\frac{\frac{1}{2}\epsilon^{ijk}\epsilon_{abc}E^b_jE^c_k}{\textrm{sgn}(\det{(E)})\sqrt{|\det{(E)}|}}.
 \end{equation}
 In fact, these two expressions of co-triad imply an alternative flux operator $\hat{E}^{i}(S)$ \cite{giesel2006consistencyi}. The consistency checking of this flux with the standard flux indicated a suitable volume operator and fixed its regularization ambiguity \cite{yang2019consistency}.  Thus it is also reasonable to consider alternative ways in the construction of the general geometric operators in all dimensional LQG, which are based on two different expressions of  the de-densitized dual momentum $\sqrt{q}\pi_{a}^{IJ}$. In this section, we will discuss how to construct general geometric operators based on the expression \eqref{qpi1}, which is similar to that of Thiemann's length operator in standard (1+3)-dimensional LQG.

Let $e_\epsilon$ be a small segment of a curve $e$ with coordinate length $\epsilon$. The de-densitized momentum can be smeared over $e_\epsilon$ as
 \begin{equation}\label{pi1}
\underline{\pi}(e_\epsilon):=\int_{e_{\epsilon}}\sqrt{q}\pi_{a}^{IJ}\tau_{IJ}\dot{e}_\epsilon^ads \hat{=}-\frac{(D-1)}{\beta\kappa}h_{e_{\epsilon}}\{h^{-1}_{e_{\epsilon}},V(v,\square)\},
 \end{equation}
 where $\tau_{IJ}$ is the basis of Lie algrbra $so(D+1)$, $h_{e_{\epsilon}}$ denotes the holonomy of the connection $A_{a}^{IJ}$ along $e_\epsilon$, $v$ is the starting point of $e_\epsilon$, and $s$ is the parameter of $e_\epsilon$. This smeared quantity can be quantized directly as,
 \begin{equation}\label{hatpi}
\hat{ \underline{\pi}}(e_\epsilon): =-\frac{(D-1)}{\mathbf{i}\beta\kappa\hbar}h_{e_{\epsilon}}[h^{-1}_{e_{\epsilon}},\hat{V}(v,\square)],
 \end{equation}
 which is called smeared de-densitized dual momentum operator. It will be used as building blocks to construct general geometric operators in all dimensional LQG.
 \subsection{The first length operator in all dimensional LQG}
  Classically, the length of a curve $e$ reads,
\begin{equation}
L_{e}=\int_{e}ds\sqrt{q_{ab}\dot{e}^a\dot{e}^b(s)},
\end{equation}
where we used equation \eqref{qab}.
Partitioning of the curve $e$ as a composition of $T$
segments
$\{e^\epsilon_t, t\in \mathbb{N}, 0\leq t\leq T\}$,
 i.e.,
\begin{equation}
 e=e^\epsilon_1\circ e^\epsilon_2\circ...\circ e^\epsilon_t\circ...\circ e^\epsilon_T,
\end{equation}
wherein $\circ$ is a composition of composable curves which
can be carried out with
\begin{equation}
e^\epsilon_t:\ [(t-1)\epsilon,t\epsilon]\rightarrow \sigma;\quad s_t\mapsto e^\epsilon_t(s_t),
\end{equation}
and $\epsilon=\frac{1}{T}$. Then, we have
\begin{equation}\label{Le=}
L_{e}=\lim_{\epsilon\rightarrow 0}\sum_{t=1}^{T}L_{e_t^\epsilon}
\end{equation}
where one has up to $o(\epsilon)$, $L_{e_t^\epsilon}\hat{=}\sqrt{\frac{1}{2} (\int_{e^\epsilon_t}ds\sqrt{q}\dot{e}^a\pi_{aIJ })(\int_{e^\epsilon_t}ds\sqrt{q}\dot{e}^b\pi_{b}^{IJ})}$. Then our task turns to be constructing the length operator $\hat{L}_{e_t^\epsilon}$ of a small curve $e_t^\epsilon$.
 By equation \eqref{qpi1} one has
\begin{eqnarray}{\label{metric1}}
q_{ab}(x)
&\hat{=}&\frac{-(D-1)^2}{2(\beta\kappa)^2}\textrm{tr}(\tau_{IJ}\tau_{KL})\{A_a^{IJ}(x),V(x,\square)\}\{A_{b}^{KL}(x),V(x,\square)\}.
\end{eqnarray}
It is easy to see that in the limit $\epsilon\rightarrow0$, we have
 \begin{equation}
 L_{e_{\epsilon}}\hat{=}\sqrt{-\frac{1}{2}\textrm{tr}(\underline{\pi}(e_\epsilon)\underline{\pi}(e_\epsilon))}.
\end{equation}
Hence, by equation \eqref{hatpi}, $L(e_{\epsilon})$ can be quantized as
\begin{eqnarray}\label{length1}
&&\hat{ L}_{e_{\epsilon}}=\sqrt{-\frac{1}{2}\textrm{tr}(\underline{\hat{\pi}}(e_\epsilon)\underline{\hat{\pi}}(e_\epsilon))} =\frac{(D-1)}{\sqrt{2}\beta\kappa\hbar}\sqrt{ \textrm{tr}(h_{e_{\epsilon}}[h^{-1}_{e_{\epsilon}},\hat{V}(v,\square)] h_{e_{\epsilon}}[h^{-1}_{e_{\epsilon}},\hat{V}(v,\square)])}\\\nonumber
&=&\frac{(D-1)}{\sqrt{2}\beta\kappa\hbar}\sqrt{(D+1)\hat{V}^2(v,\square)- \textrm{tr}(h_{e_{\epsilon}}\hat{V}(v,\square)h^{-1}_{e_{\epsilon}}\hat{V}(v,\square)) -\textrm{tr}(\hat{V}(v,\square)h_{e_{\epsilon}}\hat{V}(v,\square)h^{-1}_{e_{\epsilon}}) +\textrm{tr}(h_{e_{\epsilon}}\hat{V}^2(v,\square)h^{-1}_{e_{\epsilon}})}.
\end{eqnarray}
Denoting $\hat{\ell}_{e_{\epsilon}}:=\mathbb{I}\hat{V}(v,\square)- h_{e_{\epsilon}}\hat{V}(v,\square)h^{-1}_{e_{\epsilon}}$, we have
 \begin{equation}\label{Leepsilon}
\hat{ L}(e_{\epsilon})=\frac{(D-1)}{\sqrt{2}\beta\kappa\hbar}\sqrt{\textrm{tr}(\hat{\ell}_{e_{\epsilon}}\hat{\ell}_{e_{\epsilon}})}.
 \end{equation}
 Note that $\hat{\ell}_{e_{\epsilon}}=\hat{\ell}^\dagger_{e_{\epsilon}}$ because of $h_{e_{\epsilon}}^\dagger=h_{e_{\epsilon}}^{-1}$ and $\hat{V}(v,\square)=\hat{V}^\dagger(v,\square)$. Therefore $\hat{ L}_{e_{\epsilon}}$ is a positive
 and symmetric operator. Then the length operator for the curve $e$ can be defined as
  \begin{equation}\label{Le}
\hat{L}_{e}=\lim_{\epsilon\rightarrow 0}\sum_{t=1}^{T}\hat{L}_{e_t^\epsilon}.
  \end{equation}
  Note that, although the expression of $\hat{L}_{e}$ contains the summation of infinite terms at the limit $\epsilon\rightarrow 0$, only a finite number of terms are non-vanishing when it acts on a cylindrical state $f_\gamma$ since the volume operator only acts on nontrivial vertices $v$ of $\gamma$.  Thus the regulator $\epsilon$ can be removed in the graph-dependent manner. In the rest of the paper, all the limit of this kind of infinite summation of operators can be understood in this way. The domain of $\hat{ L}_{e}$ is the Dth order differentiable cylindrical functions satisfying the simplicity constraint in $\mathcal{H}_{\text{kin}}$, and the demonstration of its cylindrical consistency is similar to that in the standard (1+3)-dimensional LQG \cite{thiemann1998length}.

\subsection{2-Area operator in all dimensional LQG}
Although the area operator of 2-surface can be defined by the flux operator naturally in the standard (1+3)-dimensional LQG, the construction can not be directly extended to higher dimensional cases. In order to construct a 2-area operator in all dimensional case, let us come back to the (1+3)-dimensional theory to construct an alternative 2-area operator.

 We first consider the alternative flux $E_{(\textrm{alt})}^i({}^2\!S_{\diamondsuit_{12}}):=\int_{{}^2\!S}d({}^2\!S) E_{(\textrm{alt})}^{ai}n_a({}^2\!S_{\diamondsuit_{12}})$ \cite{giesel2006consistencyi} \cite{yang2019consistency}, where $E_{(\textrm{alt})}^{ai}:=\frac{1}{2}{\epsilon^i}_{jk}\epsilon^{abc}e_b^j\mathcal{S}e_c^k$, $n_a({}^2\!S_{\diamondsuit_{12}})=\frac{1}{2}\epsilon_{abc}\dot{e}^b_\imath\dot{e}^c_\jmath\epsilon^{\imath\jmath}$ with $\imath, \jmath, \imath', \jmath',=1,2$, and $\mathcal{S}:=\textrm{sgn}(\textrm{det{(e)}})$. Here $e^\epsilon_{1}, e^\epsilon_{2}$ are two linearly independent segments beginning at $v$ with coordinate length $\epsilon$, $\dot{e}^b_{1}$, $\dot{e}^c_{2}$ are their tangent vectors respectively, and ${}^2\!S_{\diamondsuit_{12}}$ is a proper open 2-surface with coordinate area $\epsilon^2$ and containing $e^\epsilon_{1}, e^\epsilon_{2}$ and $v$. Then up to $\mathcal{O}(\epsilon^2)$, we have
\begin{eqnarray}\label{altflux}
E^{(\textrm{alt})}_i({}^2\!S_{\diamondsuit_{12}})
&=&\frac{\epsilon^2}{-2\times2}\mathcal{S}\textrm{tr}(\tau_i\tau_j\tau_k)e_b^je_c^k\epsilon^{\imath\jmath} \dot{e}_\imath^{b}\dot{e}_\jmath^{c}\\\nonumber
&=&-\frac{1}{(\kappa\gamma)^2}\mathcal{S}\textrm{tr}(\tau_ih_{e^\epsilon_\imath}\{h^{-1}_{e^\epsilon_\imath},V(x,\square)\} h_{e^\epsilon_\jmath}\{h^{-1}_{e^\epsilon_\jmath},V(x,\square)\})\epsilon^{\imath\jmath}\\\nonumber
&=&-\frac{1}{(\kappa\gamma)^2}\textrm{tr}(\tau_i h_{e^\epsilon_\jmath}\{h^{-1}_{e^\epsilon_\imath},V(x,\square)\} \mathcal{S}\{h^{-1}_{e^\epsilon_\jmath},V(x,\square)\}h_{e^\epsilon_\imath})\epsilon^{\imath\jmath}
.\end{eqnarray}
Notice that in the third step of Eq.\eqref{altflux}, the ordering of holonomies $h_{e^\epsilon_\imath}$ and $h_{e^\epsilon_\jmath}$ was changed, while the contraction of their indices was kept unchanged.
Classically, it is easy to see that $E^{ai}_{(\textrm{alt})}=E^{ai}$, or $E^{(\textrm{alt})}_i({}^2\!S_{\diamondsuit_{12}})=E_i({}^2\!S_{\diamondsuit_{12}})$. Thus we can define the alternative regulated flux operator by
\begin{eqnarray}
\hat{E}^{(\textrm{alt})}_i({}^2\!S_{\diamondsuit_{12}}) &=& \frac{1}{(\kappa\gamma\hbar)^2}\textrm{tr}(\tau_i h_{e^\epsilon_\jmath}[h^{-1}_{e^\epsilon_\imath},\hat{V}(x,\square)] \mathcal{\hat{S}}[h^{-1}_{e^\epsilon_\jmath},\hat{V}(x,\square)]h_{e^\epsilon_\imath})\epsilon^{\imath\jmath}.
\end{eqnarray}
 Then a corresponding symmetric operator can be defined as
\begin{eqnarray}
\hat{E}^{\textrm{alt}}_i({}^2\!S_{\diamondsuit_{12}}) &=& \frac{1}{2}(\hat{E}^{(\textrm{alt})}_i({}^2\!S_{\diamondsuit_{12}}) +{\hat{E}^{(\textrm{alt})\dagger}_i}({}^2\!S_{\diamondsuit_{12}}) ),
\end{eqnarray}
where we ordered all the variables following the scheme in \cite{giesel2006consistencyi}\cite{yang2019consistency}. This ordering ensures that $\hat{E}^{\textrm{alt}}_i({}^2\!S_{\diamondsuit_{12}})$ is consistent with the standard flux operator $\hat{E}^i({}^2\!S_{\diamondsuit_{12}})$ at least in certain case. Now, the classical identity  $\textrm{Ar}({}^2\!S_{\diamondsuit_{12}})\approx\sqrt{E^i({}^2\!S_{\diamondsuit_{12}}) E^j({}^2\!S_{\diamondsuit_{12}})\delta_{ij}}$ indicates that we can define an alternative area operator by
  \begin{equation}\label{alterarea}
\widehat{\textrm{Ar}}_{\textrm{alt}}({}^2\!S_{\diamondsuit_{12}}) =\sqrt{\hat{E}^{\textrm{alt}}_i({}^2\!S_{\diamondsuit_{12}}) \hat{E}^{\textrm{alt}}_j({}^2\!S_{\diamondsuit_{12}})\delta^{ij}}.
  \end{equation}
  The alternative area operator can also be understood in another perspective of geometry. The classical corresponding expression of $\widehat{\textrm{Ar}}_{\textrm{alt}}({}^2\!S_{\diamondsuit_{12}})$ can be written as
 \begin{equation}
 \textrm{Ar}({}^2\!S_{\diamondsuit_{12}})\approx\sqrt{\delta_{ij}E_{\textrm{alt}}^{j}({}^2\!S_{\diamondsuit_{\imath\jmath}}) E_{\textrm{alt}}^{i}({}^2\!S_{\diamondsuit_{12}})}\approx\sqrt{\frac{\epsilon^4}{2}q_{bd}q_{ce} \epsilon^{\imath\jmath}\dot{e}_\imath^{b}\dot{e}_\jmath^{c} \epsilon^{\imath'\jmath'}\dot{e}_{\imath'}^{d}\dot{e}_{\jmath'}^{e}}.
 \end{equation}
Hence we have
 \begin{equation}
\textrm{Ar}({}^2\!S_{\diamondsuit_{\imath\jmath}})
\approx\sqrt{L^2(e^\epsilon_{1})L^2(e^\epsilon_{2})-(\epsilon\dot{e}^a_{1}(x) q_{ab}\epsilon\dot{e}^b_{2}(x))^2},
 \end{equation}
 where $L(e^\epsilon_{\imath}):=\int_{e^\epsilon_{\imath}} ds(e^\epsilon_{\imath})\sqrt{q_{ab}\dot{e}_{\imath}^{a}\dot{e}_{\imath}^{b}} \approx\epsilon\sqrt{q_{ab}\dot{e}_{\imath}^{a}\dot{e}_{\imath}^{b}(v)}$. Note that we also have
 \begin{equation}\label{costheta12}
\cos\theta_{12}=\lim_{\epsilon\rightarrow0}\frac{\epsilon^2\dot{e}^a_{1}(v) q_{ab}\dot{e}^b_{2}(v)}{L(e^\epsilon_{1})L(e^\epsilon_{2})},
 \end{equation}
  where $\theta_{12}$ is the angle between $\dot{e}^a_{1}(v) $ and $\dot{e}^b_{2}(v)$.  Therefore we obtain
 \begin{equation}
 \textrm{Ar}({}^2\!S_{\diamondsuit_{12}})
 \approx L(e^\epsilon_{1})L(e^\epsilon_{2})\sin\theta_{12},
 \end{equation}
 which is the standard expression of the area of ${}^2\!S_{\diamondsuit_{\imath\jmath}}$ in Euclidean space.

The advantage of the alternative 2-area operator \eqref{alterarea} is that its construction can be extended to all dimensional case naturally. Let us define an alternative ``flux'' operator suitable to (1+D)-dimensional cases as
\begin{eqnarray}
\hat{\tilde{E}}^{\bar{M}}_{(\textrm{gen})}({}^2\!S_{\diamondsuit_{12}})&\equiv&\hat{\tilde{E}}^{I_1...I_{D-1}}_{(\textrm{gen})}({}^2\!S_{\diamondsuit_{12}})\\\nonumber
&=&\frac{\mathrm{C}}{(\kappa\beta\hbar)^2}\epsilon^{I_1...I_{D-1}KL}\textrm{tr}^{e_1,e_2}\sum_{\imath,\jmath}(\tau^{e_\imath}_{KI}{\tau^{e_\jmath I}}_{L}\\\nonumber &&h_{e^\epsilon_\jmath}[h^{-1}_{e^\epsilon_\imath},\hat{V}(v,\square)] \hat{\mathcal{S}}[h^{-1}_{e^\epsilon_\jmath},\hat{V}(v,\square)]h_{e^\epsilon_\imath})\epsilon^{\imath\jmath},
\end{eqnarray}
where $\mathrm{C}=\frac{(D-1)^2}{2\sqrt{(D-1)!}}$, $\bar{M}$ is $(D-1)$-tuple totally asymmetric indices, $\epsilon^{I_1...I_{D+1}}$ is the Levi-Civita symbols in the internal space, $\textrm{tr}^{e_1,e_2}$ represents tracing the indices of $\tau^{e_1}_{I_{D-2}I}, h_{e^\epsilon_1}, h^{-1}_{e^\epsilon_1}$ and $\tau^{e_2}_{I_{D-2}I}, h_{e^\epsilon_2}, h^{-1}_{e^\epsilon_2}$ separately, and $\mathcal{S}$ takes value of $1$ if $(D+1)$ is odd while takes value of  1 with a sign of $\det(\pi):=\frac{1}{2D!}\epsilon_{aa_1b_1...a_nb_n}\epsilon_{IJI_1J_1...I_nJ_n}\pi^{aIJ} \pi^{a_1I_1K_1}{\pi^{b_1J_1}}_{K_1}...\pi^{a_nI_nK_n}{\pi^{b_nJ_n}}_{K_n}$ if $(D+1)$ is even. Here $\epsilon_{a_1...a_D}$ is the Levi-Civita symbol in the external space. Then the area $\textrm{Ar}({}^2\!S_{\diamondsuit_{12}})$ of the 2-surface ${}^2\!S_{\diamondsuit_{12}}$ can be promoted as an operator in all dimensional case by
\begin{equation}
\widehat{\textrm{Ar}({}^2\!S_{\diamondsuit_{12}})}=\sqrt{\hat{\tilde{E}}_{\bar{M}}^{\textrm{gen}}({}^2\!S_{\diamondsuit_{12}}) \hat{\tilde{E}}^{\bar{M}}_{\textrm{gen}}({}^2\!S_{\diamondsuit_{12}})},
\end{equation}
where
\begin{equation}
\hat{\tilde{E}}^{\bar{M}}_{\textrm{gen}}({}^2\!S_{\diamondsuit_{12}})= \frac{1}{2}(\hat{\tilde{E}}^{\bar{M}}_{(\textrm{gen})}({}^2\!S_{\diamondsuit_{12}}) +{\hat{\tilde{E}}^{\bar{M}}_{(\textrm{gen})}({}^2\!S_{\diamondsuit_{12}})}^\dag).
\end{equation}
Now we need to show that the classical analogue of $\widehat{\textrm{Ar}({}^2\!S_{\diamondsuit_{12}})}$ is exactly the 2-area. Note that up to $\mathcal{O}(\epsilon^2)$ the classical analogue of
  $\hat{\tilde{E}}^{\bar{M}}_{(\textrm{gen})}({}^2\!S_{\diamondsuit_{12}})$ reads
\begin{eqnarray}
\tilde{E}^{\bar{M}}_{(\textrm{gen})}({}^2\!S_{\diamondsuit_{12}})
&=&-\frac{\mathrm{C}}{(\kappa\beta)^2}\epsilon^{I_1...I_{D-1}KL}\textrm{tr}^{e_1,e_2}\sum_{\imath,\jmath}(\tau^{e_\imath}_{KI}{\tau^{e_\jmath I}}_{L}\\\nonumber &&h_{e^\epsilon_\jmath}\{h^{-1}_{e^\epsilon_\imath},V(v,\square)\} \mathcal{S}\{h^{-1}_{e^\epsilon_\jmath},V(v,\square)\}h_{e^\epsilon_\imath})\epsilon^{\imath\jmath}\\\nonumber
&=&\frac{\mathrm{C}}{(\kappa\beta)^2}\epsilon^2\epsilon^{I_1...I_{D-1}KL}\textrm{tr}(\tau_{KI}\tau^{J_1J'_1})\textrm{tr}({\tau_{L} }^I\tau^{J_2J'_2})\\\nonumber &&\{A_{aJ_1J'_1}\dot{e}^a_{\imath},V(v,\square)\} \mathcal{S}\{A_{bJ_2J'_2}\dot{e}^b_{\jmath},V(v,\square)\})\epsilon^{\imath\jmath}\\\nonumber
&=&-\frac{\mathrm{C}}{(D-1)^2}\epsilon^2\epsilon^{I_1...I_{D-1}KL} \dot{e}^a_{\imath}\sqrt{q}{\pi_{aK}}^{I} \mathcal{S}\dot{e}^b_{\jmath}{\sqrt{q}\pi_{b}}_{IL}\epsilon^{\imath\jmath}.
\end{eqnarray}
Hence, up to $\mathcal{O}(\epsilon^4)$, we have
\begin{eqnarray}
\delta_{\bar{M}\bar{M'}}\tilde{E}^{\bar{M}}_{(\textrm{gen})}\tilde{E}^{\bar{M'}}_{(\textrm{gen})}({}^2\!S_{\diamondsuit_{12}}) &=&\frac{\mathrm{C}^2}{(D-1)^4}\epsilon^4(D-1)!2!\delta^{[K}_{K'}\delta^{L]}_{L'} \\\nonumber &&\dot{e}^a_{\imath}{\sqrt{q}\pi_{aK}}^{I} \dot{e}^b_{\jmath}{\sqrt{q}\pi_{b}}_{IL}\epsilon^{\imath\jmath} \dot{e}^{a'}_{\imath'}\sqrt{q}{\pi^{K'}_{a'I'}} \dot{e}^{b'}_{\jmath'}{\sqrt{q}\pi_{b'}}^{I'L'}\epsilon^{\imath'\jmath'}\\\nonumber
&=&\frac{1}{2}\epsilon^4q_{aa'}q_{bb'}\dot{e}^a_{\imath} \dot{e}^b_{\jmath}\epsilon^{\imath\jmath} \dot{e}^{a'}_{\imath'} \dot{e}^{b'}_{\jmath'}\epsilon^{\imath'\jmath'}\\\nonumber
&=&(L(e^\epsilon_{1})L(e^\epsilon_{2})\sin\theta_{12})^2.
\end{eqnarray}
Therefore, the classical analogue of $\widehat{\textrm{Ar}({}^2\!S_{\diamondsuit_{12}})}$ does correspond to the classical 2-area expression $L(e^\epsilon_{1})L(e^\epsilon_{2})\sin\theta_{12}$.

 In the above construction of $\widehat{\textrm{Ar}({}^2\!S_{\diamondsuit_{12}})}$, we generalized the alternative 2-area operator in the standard (1+3)-dimensional LQG to higher dimensional case, and kept the authentic ordering of its constituents, which has been shown to be consistent with the standard flux operator in certain case.

\subsection{General $m$-area operators in all dimensional LQG}
The above construction of the 2-area operator inspires us to consider more general case. We will use a similar way to construct $m$-dimensional ($1\leq m\leq D$) area operators in (1+D)-dimensional LQG. Consider a partition ${}^m\!S=\bigcup_{t=1}^T {}^m\!\overline{S}^t_{\diamondsuit_{1m}}$ of an open $m$-surface ${}^m\!S$, with ${}^m\!\overline{S}^t_{\diamondsuit_{1m}}$ being closed  $m$-surface with open interior ${}^m\!S^t_{\diamondsuit_{1m}}$, $t$ being the labeling number of these component $m$-surfaces in this partition, and $T$ being the total number of them.
 The arbitrary small $m$-surfaces ${}^m\!S_{\diamondsuit_{1m}}$ has coordinate area $\epsilon^m$ and contains $e^\epsilon_1, e^\epsilon_2,..., e^\epsilon_m$ and $v$, where the $m$ small segments $e^\epsilon_1, e^\epsilon_2,..., e^\epsilon_m$ have common beginning point $v$ and coordinate length $\epsilon$. Their tangent vectors $\dot{e}^a_1(v), \dot{e}^a_2(v),..., \dot{e}^a_m(v)$ at point $v$ span a $m$-dimensional vector space. A local right-handed coordinate system $\{s_1,...,s_m\}$ can be  defined such that $v=(0,...,0)$, $t(e^\epsilon_\imath)=(0,...,s_\imath,...,0)|_{s_\imath=\epsilon}$, $\dot{e}^a_\imath=(\frac{\partial}{\partial s_\imath})^a|_{e^\epsilon_\imath}$, and $e^\epsilon_1, e^\epsilon_2,..., e^\epsilon_m$ are its positive oriented coordinate axis, where $\imath=1,...,m$, and $t(e^\epsilon_\imath)$ is the target point of $e^\epsilon_\imath$.
The classical expression of  the $m$-area of ${}^m\!S$ reads
\begin{equation}
\textrm{Ar}({}^m\!S)=\sum\textrm{Ar}({}^m\!S_{\diamondsuit_{1m}})=\sum\int_{{}^m\!S_{\diamondsuit_{1m}}}\sqrt{\det({}^m\!q)}ds_1...ds_m,
\end{equation}
where
\begin{equation}
\det({}^m\!q)=\frac{1}{m!}{}^m\!q_{a_1a'_1}...{}^m\!q_{a_ma'_m} \dot{e}^{a_1}_{\imath_1}...\dot{e}^{a_m}_{\imath_m}\epsilon^{\imath_1...\imath_m} \dot{e}^{a'_1}_{\imath'_1}...\dot{e}^{a'_m}_{\imath'_m}\epsilon^{\imath'_1...\imath'_m}
\end{equation}
denotes the determinant of the metric ${}^m\!q_{ab}$ on ${}^m\!S_{\diamondsuit_{1m}}$ induced by $q_{ab}$. To construct the $m$-area operator, we need to consider the following two cases separately.
 \subsubsection{Case I: $\mathbf{m}$ is even}
Taking account of the identity \eqref{qab}, we define the $m$-form component
\begin{eqnarray}\label{Em}
\tilde{E}_{(\textrm{gen})}^{I_1...I_m}&:=&\frac{1}{\sqrt{m!}}\mathcal{S}\sqrt{q}\pi_{a_1}^{I_1J_1}\delta_{J_1J_2}\sqrt{q}\pi_{a_2}^{I_2J_2} ...\\\nonumber &&\sqrt{q}\pi_{a_{m-1}}^{I_{m-1}J_{m-1}}\delta_{J_{m-1}J_m}\sqrt{q}\pi_{a_m}^{I_mJ_m}\dot{e}^{a_1}_{\imath_1}...\dot{e}^{a_m}_{\imath_m}\epsilon^{\imath_1...\imath_m}
\end{eqnarray}
such that
\begin{equation}
\det({}^m\!q)\hat{=}\tilde{E}_{(\textrm{gen})}^{I_1...I_m}\tilde{E}_{(\textrm{gen})I_1...I_m}.
\end{equation}
The general flux of $\tilde{E}_{(\textrm{gen})}^{I_1...I_m}$ is defined as
\begin{equation}
\tilde{E}_{(\textrm{gen})}^{I_1...I_m}({}^m\!S_{\diamondsuit_{1m}}):= \int_{e^\epsilon_1}...\int_{e^\epsilon_m}\tilde{E}_{(\textrm{gen})}^{I_1...I_m}ds_1...ds_m.
\end{equation}
Then, up to $\mathcal{O}(\epsilon^m)$ we have
\begin{eqnarray}
&&\tilde{E}_{(\textrm{gen})}^{I_1...I_m}({}^m\!S_{\diamondsuit_{1m}})\\\nonumber
 &=&\frac{(-1)^m}{\sqrt{m!}}\int_{e^\epsilon_1}...\int_{e^\epsilon_m}\textrm{tr}(\tau^{I_1J_1}\tau_{I''_1J''_1}) \mathcal{S} \sqrt{q}\pi_{a_1}^{I''_1J''_1}\delta_{J_1J_2}\textrm{tr}(\tau^{I_2J_2}\tau_{I''_2J''_2}) \sqrt{q}\pi_{a_2}^{I''_2J''_2}\\\nonumber
 &&...\delta_{J_{m-1}J_m}\textrm{tr}(\tau^{I_mJ_m}\tau_{I''_mJ''_m}) \sqrt{q}\pi_{a_m}^{I''_mJ''_m}\dot{e}^{a_1}_{\imath_1}...\dot{e}^{a_m}_{\imath_m}\epsilon^{\imath_1...\imath_m}ds_1...ds_m\\\nonumber
 &\hat{=}&(\frac{(D-1)^m}{(\beta\kappa)^m\sqrt{m!}}) \textrm{tr}^{e^\epsilon_1...e^\epsilon_m}(\tau_{e^\epsilon_{\imath_1}}^{I_1J_1}\delta_{J_1J_2}\tau_{e^\epsilon_{\imath_2}}^{I_2J_2} ...\delta_{J_{m-1}J_m}\tau_{e^\epsilon_{\imath_m}}^{I_mJ_m} \\\nonumber
 &&\mathcal{S}h_{e_{\imath_1}^\epsilon}\{h_{e_{\imath_1}^\epsilon}^{-1},V(v,\square)\} ...h_{e_{\imath_m}^\epsilon}\{h_{e_{\imath_m}^\epsilon}^{-1},V(v,\square)\})\epsilon^{\imath_1...\imath_m}.
\end{eqnarray}
Inspired by the ordering of the alternative 2-area operator, the general flux can be quantized as
\begin{eqnarray}
&&\widehat{\tilde{E}_{(\textrm{gen})}^{I_1...I_m}}({}^m\!S_{\diamondsuit_{1m}})\\\nonumber
&=& (\frac{(D-1)^m}{(\mathbf{i}\beta\kappa\hbar)^m\sqrt{m!}}) \textrm{tr}^{e^\epsilon_1...e^\epsilon_m}(\tau_{e^\epsilon_{\imath_1}}^{I_1J_1}\delta_{J_1J_2}\tau_{e^\epsilon_{\imath_2}}^{I_2J_2}...\delta_{J_{m-1}J_m}\tau_{e^\epsilon_{\imath_m}}^{I_mJ_m} \widehat{\textrm{Per.}})\epsilon^{\imath_1...\imath_m}
,\quad\epsilon\mapsto0,
\end{eqnarray}
where
\begin{eqnarray}\label{Per}
\widehat{\textrm{Per.}}&\equiv&h_{e^\epsilon_1}h_{e^\epsilon_2} ...h_{e^\epsilon_{m-1}}h_{e^\epsilon_m}\hat{V}_{\textrm{tot}}(v,\square)h_{e^\epsilon_1}^{-1}h_{e^\epsilon_2}^{-1} ...h_{e^\epsilon_{m-1}}^{-1}h_{e^\epsilon_m}^{-1}.
\end{eqnarray}
 Here we denote
  \begin{equation}\label{Vtot}
\hat{V}_{\textrm{tot}}(v,\square)=\sum_{\iota_1,...,\iota_m}(-1)^{\vartheta(\iota_1,...,\iota_m)} \hat{V}_{\iota_11}(v,\square)\hat{V}_{\iota_22}(v,\square)...\hat{V}_{\iota_{[m/2]}, [m/2]}(v,\square)\widehat{\mathcal{S}}\hat{V}_{\iota_{[m/2]+1},[m/2]+1}(v,\square)...\hat{V}_{\iota_m m}(v,\square),
  \end{equation}
  where $\iota_1,...,\iota_m$ takes values from  $+$ and $-$,  $\vartheta(\iota_1,...,\iota_m)$ is the total numbers of $-$ in $\{\iota_1,...,\iota_m\}$, and thus $\hat{V}_{\iota\imath}(v,\square)$ denotes one of the elements of the matrix
\begin{equation}
\left(
\begin{array}{ccccc}
\hat{V}_{+1}(v,\square)&\hat{V}_{+2}(v,\square)&...&\hat{V}_{+(m-1)}(v,\square)&\hat{V}_{+m}(v,\square)\\
\hat{V}_{-1}(v,\square)&\hat{V}_{-2}(v,\square)&...&\hat{V}_{-(m-1)}(v,\square)&\hat{V}_{-m}(v,\square)
\end{array}
\right)
\end{equation}
Our requirement is that $\hat{V}_{+\imath}(v,\square)$ acts on the holonomies except $h_{e^\epsilon_{\imath}}^{-1}$ on its right, while $\hat{V}_{-\imath}(v,\square)$ acts on all the holonomies on its right. The symmetric version of the general flux operator reads
\begin{equation}
\widehat{\tilde{E}_{\textrm{gen}}^{I_1...I_m}}({}^m\!S_{\diamondsuit_{1m}})=\frac{1}{2}(\widehat{\tilde{E}_{(\textrm{gen})}^{I_1...I_m}}({}^m\!S_{\diamondsuit_{1m}}) +\widehat{{\tilde{E}}_{(\textrm{gen})}^{I_1...I_m}}^\dagger({}^m\!S_{\diamondsuit_{1m}})).
\end{equation}
Since classically one has
\begin{equation}
\textrm{Ar}({}^m\!S)=\lim_{\epsilon\rightarrow0}\sum_{{}^m\!S_{\diamondsuit_{1m}}}\textrm{Ar}({}^m\!S_{\diamondsuit_{1m}}) =\lim_{\epsilon\rightarrow0}\sum_{{}^m\!S_{\diamondsuit_{1m}}}\sqrt{\tilde{E}_{(\textrm{gen})} ^{I_1...I_m}({}^m\!S_{\diamondsuit_{1m}})\tilde{E}_{(\textrm{gen})I_1...I_m}({}^m\!S_{\diamondsuit_{1m}})},
\end{equation}
 we propose the $m$-area operator as
\begin{equation}
\widehat{\textrm{Ar}({}^m\!S)} =\lim_{\epsilon\rightarrow0}\sum_{{}^m\!S_{\diamondsuit_{1m}}}\widehat{\textrm{Ar}({}^m\!S_{\diamondsuit_{1m}})} =\lim_{\epsilon\rightarrow0}\sum_{{}^m\!S_{\diamondsuit_{1m}}} \sqrt{\widehat{\tilde{E}_{\textrm{gen}}^{I_1...I_m}}({}^m\!S_{\diamondsuit_{1m}}) \widehat{\tilde{E}^{\textrm{gen}}_{I_1...I_m}}({}^m\!S_{\diamondsuit_{1m}})}.
\end{equation}
Similar to the construction of the the length operator $\hat{L}_e$, the regulator $\epsilon$ can be removed in the graph-dependent way.
 \subsubsection{Case II: $\mathbf{m}$ is odd}
 Similar to the case I, classically we define the $m$-form component
 \begin{eqnarray}\label{Em+1}
\tilde{E}_{(\textrm{gen})}^{II_1...I_m}&:=&\frac{1}{\sqrt{2m!}}\mathcal{S}\sqrt{q}\pi_{a_1}^{II_1}\sqrt{q}\pi_{a_2}^{I_2J_2}\delta_{J_2J_3}\sqrt{q}\pi_{a_3}^{I_3J_3} ...\\\nonumber &&\sqrt{q}\pi_{a_{m-1}}^{I_{m-1}J_{m-1}}\delta_{J_{m-1}J_m}\sqrt{q}\pi_{a_m}^{I_mJ_m}\dot{e}^{a_1}_{\imath_1}...\dot{e}^{a_m}_{\imath_m}\epsilon^{\imath_1...\imath_m}.
\end{eqnarray}
Note that there are $m+1$ internal indices in \eqref{Em+1}, while \eqref{Em} contains only $m$ internal indices. Then it is easy to see that
\begin{equation}
\det({}^m\!q)\hat{=}\tilde{E}_{(\textrm{gen})}^{II_1...I_m}\tilde{E}_{(\textrm{gen})II_1...I_m}.
\end{equation}
The generalized flux of the $m$-form can be expressed up to $\mathcal{O}(\epsilon^m)$ as
\begin{eqnarray}
\tilde{E}_{(\textrm{gen})}^{II_1...I_m}({}^m\!S_{\diamondsuit_{1m}})
&:=& \int_{e^\epsilon_1}...\int_{e^\epsilon_m}\tilde{E}_{(\textrm{gen})}^{II_1...I_m}ds_1...ds_m\\\nonumber
 &=&\frac{(-1)^m}{\sqrt{2m!}}\int_{e^\epsilon_1}...\int_{e^\epsilon_m}\textrm{tr}(\tau^{II_1}\tau_{I''I''_1}) \mathcal{S}\sqrt{q}\pi_{a_1}^{I''I''_1}\textrm{tr}(\tau^{I_2J_2}\tau_{I''_2J''_2}) \sqrt{q}\pi_{a_2}^{I''_2J''_2}\delta_{J_2J_3}\\\nonumber
 &&...\delta_{J_{m-1}J_m}\textrm{tr}(\tau^{I_mJ_m}\tau_{I''_mJ''_m}) \sqrt{q}\pi_{a_m}^{I''_mJ''_m}\dot{e}^{a_1}_{\imath_1}...\dot{e}^{a_m}_{\imath_m}\epsilon^{\imath_1...\imath_m}ds_1...ds_m\\\nonumber
 &\hat{=}&(\frac{(D-1)^m}{(\beta\kappa)^m\sqrt{2m!}}) \textrm{ tr}^{e^\epsilon_1...e^\epsilon_m}(\tau_{e^\epsilon_{\imath_1}}^{II_1}\tau_{e^\epsilon_{\imath_2}}^{I_2J_2}\delta_{J_2J_3}...\delta_{J_{m-1}J_m}\tau_{e^\epsilon_{\imath_m}}^{I_mJ_m} \\\nonumber
 &&\mathcal{S}h_{e_{\imath_1}^\epsilon}\{h_{e_{\imath_1}^\epsilon}^{-1},V(v,\square)\} ...h_{e_{\imath_m}^\epsilon}\{h_{e_{\imath_m}^\epsilon}^{-1},V(v,\square)\})\epsilon^{\imath_1...\imath_m}.
\end{eqnarray}
Following the same quantization procedures as case I, we have
\begin{eqnarray}
&&\widehat{\tilde{E}_{(\textrm{gen})}^{II_1...I_m}}({}^m\!S_{\diamondsuit_{1m}})\\\nonumber
&:=& (\frac{(D-1)^m}{(\mathbf{i}\beta\kappa\hbar)^m\sqrt{2m!}}) \textrm{tr}^{e^\epsilon_1...e^\epsilon_m}(\tau_{e^\epsilon_{\imath_1}}^{II_1}\tau_{e^\epsilon_{\imath_2}}^{I_2J_2}\delta_{J_2J_3}...\delta_{J_{m-1}J_m}\tau_{e^\epsilon_{\imath_m}}^{I_mJ_m} \widehat{\textrm{Per.}})\epsilon^{\imath_1...\imath_m},
\end{eqnarray}
where $\widehat{\textrm{Per.}}$ was defined by \eqref{Per}. The symmetric generalized flux operator can be defined by
\begin{equation}
\widehat{\tilde{E}_{\textrm{gen}}^{II_1...I_m}}({}^m\!S_{\diamondsuit_{1m}}):=\frac{1}{2} (\widehat{\tilde{E}_{(\textrm{gen})}^{II_1...I_m}}({}^m\!S_{\diamondsuit_{1m}}) +\widehat{\tilde{E}_{(\textrm{gen})}^{II_1...I_m}}^\dagger({}^m\!S_{\diamondsuit_{1m}})).
\end{equation}
Again, classically one has the area expression
\begin{equation}
\textrm{Ar}({}^m\!S)=\lim_{\epsilon\rightarrow0}\sum_{{}^m\!S_{\diamondsuit_{1m}}}\textrm{Ar}({}^m\!S_{\diamondsuit_{1m}}) =\lim_{\epsilon\rightarrow0}\sum_{{}^m\!S_{\diamondsuit_{1m}}}\sqrt{\tilde{E}_{(\textrm{gen})}^{II_1...I_m} ({}^m\!S_{\diamondsuit_{1m}})\tilde{E}_{(\textrm{gen})II_1...I_m}({}^m\!S_{\diamondsuit_{1m}})}.
\end{equation}
Hence the $m$-area operator for this case is proposed as
\begin{equation}
\widehat{\textrm{Ar}({}^m\!S)} =\lim_{\epsilon\rightarrow0}\sum_{{}^m\!S_{\diamondsuit_{1m}}}\widehat{\textrm{Ar}({}^m\!S_{\diamondsuit_{1m}})} =\lim_{\epsilon\rightarrow0}\sum_{{}^m\!S_{\diamondsuit_{1m}}}\sqrt{\widehat{\tilde{E}_{\textrm{gen}}^{II_1...I_m}}({}^m\!S_{\diamondsuit_{1m}}) \widehat{\tilde{E}^{\textrm{gen}}_{II_1...I_m}}({}^m\!S_{\diamondsuit_{1m}})}.
\end{equation}

\section{Issues of the general $m$-area operators}

 \subsection{The ambiguity in the construction of geometric operators}
Let us consider two special cases of the general $m$-area operator where an ambiguity in the construction of geometric operators will appear.
In the special case of $m=1$, the general $m$-area operator of ${}^m\!S_{\diamondsuit_{1m}}$  becomes a length operator  of $e^\epsilon$ as
 \begin{equation}\label{length2}
\widehat{L_{\textrm{alt}}({e^\epsilon})} =\sqrt{\widehat{\tilde{E}_{\textrm{gen}}^{II_1}}({e^\epsilon})\widehat{\tilde{E}^{\textrm{gen}}_{II_1}}({e^\epsilon})},
\end{equation}
where
\begin{eqnarray}
\widehat{\tilde{E}_{\textrm{gen}}^{II_1}}({e^\epsilon})
 &=&\frac{-(D-1)}{\mathbf{i}\beta\kappa\hbar\sqrt{2}}\textrm{ tr}(\tau_{e^\epsilon}^{II_1} h_{e^\epsilon}\widehat{V}(v,\square)h_{e^\epsilon}^{-1}).
\end{eqnarray}
From Eq. \eqref{length2} we get
\begin{equation}\label{length22}
\widehat{L_{\textrm{alt.}}({e^\epsilon})}  =\frac{(D-1)}{\sqrt{2}\beta\kappa\hbar}\sqrt{ {(h_{e^\epsilon})^I}_J\widehat{V}(v,\square)\widehat{V}(v,\square){(h_{e^\epsilon}^{-1})^J}_I-{(h_{e^\epsilon})_I}^J\widehat{V}(v,\square) {(h_{e^\epsilon}^{-1})_J}^K{(h_{e^\epsilon})^I}_{J'}\widehat{V}(v,\square){({h_{e^\epsilon}^{-1}})^{J'}}_K }.
\end{equation}
Recall that the generalization of Thiemann's length operator was given by Eq.\eqref{length1}, which is different from Eq.\eqref{length22} formally. In fact, this difference comes from the different choices of the ordering of the holonomies and volumes in the expressions of $q_{ab}$.

In the case of $m=D$, the  general $m$-area operator of ${}^m\!S_{\diamondsuit_{1m}}$ becomes alternative $D$-volume operators as
\begin{equation}\label{altvol1}
\widehat{\textrm{Vol}_{\textrm{alt}}}({}^D\!S_{\diamondsuit_{1D}}) =\sqrt{\widehat{\tilde{E}_{\textrm{gen}}^{II_1...I_m}}({}^D\!S_{\diamondsuit_{1D}}) \widehat{\tilde{E}^{\textrm{gen}}_{II_1...I_m}}({}^D\!S_{\diamondsuit_{1D}})}
\end{equation}
for odd $D$, and
\begin{equation}\label{altvol2}
\widehat{\textrm{Vol}_{\textrm{alt}}}({}^D\!S_{\diamondsuit_{1D}}) =\sqrt{\widehat{\tilde{E}_{\textrm{gen}}^{I_1...I_m}}({}^D\!S_{\diamondsuit_{1D}}) \widehat{\tilde{E}^{\textrm{gen}}_{I_1...I_m}}({}^D\!S_{\diamondsuit_{1D}})}
\end{equation}
for even $D$. The alternative $D$-volume operators \eqref{altvol1} and \eqref{altvol2} are totally constructed by the dual momentum operator which contains the standard $D$-volume operator. There is an analogous alternative volume operator $\widehat{\textrm{Vol}_{\textrm{alt}}^{\text{sta}}}$ in the standard (1+3)-dimensional LQG \cite{yang2016new}. Consider the case of $D=3$ and denote by $\widehat{\textrm{Vol}^{\textrm{all}}_{\text{alt}}}$ the operators \eqref{altvol1} and \eqref{altvol2} in this case. It is interesting to compare $\widehat{\textrm{Vol}^{\textrm{all}}_{\text{alt}}}$ with $\widehat{\textrm{Vol}^{\textrm{sta}}_{\text{alt}}}$. There are following two main differences between them. Firstly, the dual momentum used to construct $\widehat{\textrm{Vol}^{\textrm{all}}_{\textrm{alt}}}$ is $so(4)$-valued, while the co-triad used to construct $\widehat{\textrm{Vol}^{\textrm{sta}}_{\textrm{alt}}}$ is Lie algebra $su(2)$-valued. Secondly, the construction schemes and ingredients of the two operators are different. To construct $\widehat{\textrm{Vol}^{\textrm{sta}}_{\textrm{alt}}}$, one employed the classical identity $\textrm{Vol}^{\textrm{sta}}_{\textrm{alt}}(R):=\int_Rd^3x|\det{(e)}|$ with Eq.\eqref{stane}, where $V(v,\square)$ is quantized as the volume operator in standard (1+3)-dimensional LQG. To construct $\widehat{\textrm{Vol}^{\textrm{all}}_{\textrm{alt}}}$, we employed the classical identity $\textrm{Vol}^{\textrm{all}}_{\textrm{alt}}(R):=\int_Rd^Dx\sqrt{\det(q)}$ with Eqs.\eqref{qab} and \eqref{qpi1}, where $V(v,\square)$ is quantized as the standard volume operator in all dimensional LQG.

In fact, the ambiguity appeared in the construction of the above $m$-area operators is rather general in the construction of other geometric operators.

\subsection{The issue of simplicity constraint}

In the construction of geometric operators in all dimensional LQG, there is the issue of how to carry out the simplicity constraint. Classically, on the simplicity constraint surface of the phase space, one has $\pi^{a}_{IJ}=2\sqrt{q}n_{[I}e^{a}_{J]}$ and $\pi_a^{IJ}=2\sqrt{q}^{-1}n^{[I}e_a^{J]}$. Hence the identity \eqref{qpi1} holds. By quantization, one expects that this ``simple'' property be transformed as the requirement to the right invariant vector fields such that $\hat{N}^{[I}R_{e_\imath}^{JK]}=0$, $\forall b(e_\imath)=v$, where $\hat{N}^I$ is the operator of an auxiliary internal vector field which plays the role of $n^I$, and $b(e_i)$ denotes the beginning point of $e_i$. In the construction of a general geometric operator, usually there would appear the following term acting on a state $f_\gamma$ as
\begin{equation}\label{hhvhh}
h^{-1}_{e_\epsilon^{\imath_1}}  ...h^{-1}_{e_\epsilon^{\imath_2}}...\hat{V}^{m_1}_{\square_\epsilon}h_{e_\epsilon^{\imath_1}}...\hat{V}^{m_2}_{\square_\epsilon} h_{e_\epsilon^{\imath_2}}...f_\gamma,
\end{equation}
where $f_\gamma$ is supposed to satisfy the simplicity constraint by labelling its edges by the simple representation of $SO(D+1)$ and its vertices by the simple intertwiners. Note that the volume operator can keep its geometric meaning only on the state satisfying the simplicity constraint. However, the holonomy operator may change the simple intertwiner into non-simple one. Hence, the operator \eqref{hhvhh} already lost its geometric meaning. One possible solution to this problem is to introduce a projection operator $\widehat{\mathbb{P}}_{S}$, which projects the space of the kinematic states into the solution space of simplicity constraint, and insert it into the two sides of each volume operator in \eqref{hhvhh} to define,
\begin{equation}
 h^{-1}_{e_\epsilon^{\imath_1}}  ...h^{-1}_{e_\epsilon^{\imath_2}}...\widehat{\mathbb{P}}_{S}\hat{V}^{m_1}_{\square_\epsilon}\widehat{\mathbb{P}}_{S} h_{e_\epsilon^{\imath_1}}...\widehat{\mathbb{P}}_{S}\hat{V}^{m_2}_{\square_\epsilon}\widehat{\mathbb{P}}_{S} h_{e_\epsilon^{\imath_2}}...f_\gamma.
\end{equation}
Generally, the degrees of freedom that should be eliminated by the simplicity constraints in the construction of a geometric operator are still unclear. This issue needs further investigation. Moreover, there is the issue of anomaly for the quantum simplicity constraint \cite{bodendorfer2013newiii} \cite{bodendorfer2013implementation}. It is argued that only the weak solutions of the quantum simplicity constraints have the reasonable physical degrees of freedom \cite{Gaoping2019coherentintertwiner}. In next section, we will introduce another scheme for constructing general geometric operators, which leads to a better behaviour of the operators concerning the issue of simplicity constraints.
\subsection{Consistency of the alternative flux and the standard flux operators}

In the special case of $m=D-1$ the $m$-area operator introduced in last section is alternative to the standard $(D-1)$-area operator defined by the standard flux operator. It is worth checking whether the two versions of area operators are consistent with each other. Now we consider the case that $(D-1)$ is even. Since the alternative $(D-1)$-area operator is consist of the alternative flux operator
 \begin{equation}\label{pialt}
\hat{\pi}^{IJ}_{\textrm{alt}}({}^{(D-1)}\!S_{\diamondsuit_{(D-1)}}):=\frac{1}{\sqrt{2(D-1)!}} {\epsilon^{IJ}}_{I_1...I_{D-1}}\widehat{\tilde{E}_{\textrm{gen} }^{I_1...I_{D-1}}}({}^{(D-1)}\!S_{\diamondsuit_{(D-1)}}),
 \end{equation}
 the necessary condition for the consistency of the two versions of area operator is the consistency of $\hat{\pi}^{IJ}_{\textrm{alt}}({}^{(D-1)}\!S_{\diamondsuit_{(D-1)}})$ with the standard flux operator $\hat{\pi}^{IJ}({}^{(D-1)}\!S_{\diamondsuit_{(D-1)}})$. Now we check this issue. Note that the action of volume operator in the expression of the alternative flux on a cylindrical function $f_\gamma$ is given by
 \begin{equation}
 \hat{V}(v,\square)\cdot f_\gamma=(\hbar\kappa\beta)^{\frac{D}{D-1}}|c_{\textrm{reg.}}\frac{\mathbf{i}^D}{D!}\sum_{e_1,...,e_D\in E(\gamma),e_1\cap...\cap e_D=v}q_{e_1,...,e_D}|^{\frac{1}{D-1}}\cdot f_\gamma,
 \end{equation}
 where
 \begin{equation}
 q_{e_1,...,e_D}=\frac{1}{2}\epsilon_{IJI_1J_1I_2J_2...I_nJ_n}R_e^{IJ}R_{e_1}^{I_1K_1}R_{e'_1K_1}^{J_1}... R_{e_n}^{I_nK_n}R_{e'_nK_n}^{J_n},
 \end{equation}
 with $R_e^{IJ}:=\textrm{tr}((\tau^{IJ}h_e(A))^T\frac{\partial}{\partial h_e(A)})$.
 Let $T_{\gamma, {}^{(D-1)}\!S_{\diamondsuit_{1,(D-1)}}}$ be a spin network state intersects the surface ${}^{(D-1)}\!S_{\diamondsuit_{1,(D-1)}}$ by an inner point $v$ of its edge $e_0$. By the identity
  \begin{equation}\label{VSV}
  [\hat{V}(v,\square)]^{\frac{D-1}{2}} \widehat{\mathcal{S}}[\hat{V}(v,\square)]^{\frac{D-1}{2}}=(\mathbf{i}\hbar\kappa\beta)^{D}\frac{c_{\textrm{reg.}}}{D!}\sum_{e_1,...,e_D\in E(\gamma),e_1\cap...\cap e_D=v}q_{e_1,...,e_D},
  \end{equation}
  the action of $\widehat{\tilde{E}_{\textrm{gen}}^{I_1...I_{D-1}}}({}^{(D-1)}\!S_{\diamondsuit_{(D-1)}})$ on $T_{\gamma, {}^{(D-1)}\!S_{\diamondsuit_{1,(D-1)}}}$ reads
\begin{eqnarray}
&&\widehat{\tilde{E}_{\textrm{gen} }^{I_1...I_{D-1}}}({}^{(D-1)}\!S_{\diamondsuit_{(D-1)}})\cdot T_{\gamma, {}^{(D-1)}\!S_{\diamondsuit_{1,(D-1)}}}\\\nonumber
&=&(\mathbf{i}\hbar\kappa\beta)^D\frac{c_{\textrm{reg.}}}{D!}(\frac{(D-1)^{(D-1)}}{(\mathbf{i}\hbar\kappa\beta)^{(D-1)}\sqrt{(D-1)!}}) \textrm{tr}^{e^\epsilon_1...e^\epsilon_{(D-1)}}( \tau_{e^\epsilon_{\imath_1}}^{I_1J_1}\delta_{J_1J_2}\tau_{e^\epsilon_{\imath_2}}^{I_2J_2}...\delta_{J_{{(D-2)}}J_{(D-1)}}\tau_{e^\epsilon_{\imath_{(D-1)}}}^{I_{(D-1)}J_{(D-1)}} \\\nonumber
&&h_{e^\epsilon_1}h_{e^\epsilon_2} ...h_{e^\epsilon_{{(D-2)}}}h_{e^\epsilon_{(D-1)}}\mathbb{\widehat{P}}_S\epsilon_{I'J'I'_1J'_1...I'_nJ'_n}R_{e_0}^{I'J'}R^{I'_1K'_1}_{e_1} {R_{e'_1 K'_1}^{J'_1}}\\\nonumber
&&...R^{I'_nK'_n}_{e_n} {R_{e'_n K'_n}^{J'_n}}\mathbb{\widehat{P}}_Sh_{e^\epsilon_1}^{-1}h_{e^\epsilon_2}^{-1} ...h_{e^\epsilon_{(D-2)}}^{-1}h_{e^\epsilon_{(D-1)}}^{-1})\epsilon^{\imath_1...\imath_{(D-1)}}\cdot T_{\gamma, {}^{(D-1)}\!S_{\diamondsuit_{1,(D-1)}}}\\\nonumber
&\sim&(\mathbf{i}\hbar\kappa\beta)^D\frac{1}{D!}\frac{c_{\textrm{reg.}}(D-1)^{(D-1)}}{(\mathbf{i}\hbar\kappa\beta)^{(D-1)} \sqrt{(D-1)!}}(\frac{1}{4})^{\frac{D-1}{2}} (D-1)! R_{e_0}^{I'J'}{\epsilon_{I'J'}}^{I_1I_2...I_{D-1}}\cdot T_{\gamma, {}^{(D-1)}\!S_{\diamondsuit_{1,(D-1)}}},
\end{eqnarray}
as $\epsilon \rightarrow 0$, where $e_1=e^\epsilon_1$, $e'_1=e^\epsilon_2$, $e_2=e^\epsilon_3$, $e'_2=e^\epsilon_4$,...,$e_n=e^\epsilon_{D-2}$, $e'_n=e^\epsilon_{D-1}$, $n=\frac{D-1}{2}$, equation $\lim_{\epsilon\rightarrow0}\textrm{tr}(\tau_{IJ} h_{e^\epsilon_{\imath}}R_{e^\epsilon_{\imath}}^{KL}h_{e^\epsilon_{\imath}}^{-1})=\delta^K_{[I}\delta^L_{J]}$ was used, and the symbol $\sim$ represents ``be proportional to''. Hence we obtain
\begin{equation}
  \hat{\pi}^{IJ}_{\textrm{alt}}({}^{(D-1)}\!S_{\diamondsuit_{(D-1)}})\cdot T_{\gamma, {}^{(D-1)}\!S_{\diamondsuit_{1,(D-1)}}}\sim c_{\textrm{reg.}}\sqrt{2}\frac{\mathbf{i}\hbar\kappa\beta}{D}(\frac{D-1}{2})^{(D-1)} R_{e_0}^{IJ}\cdot T_{\gamma, {}^{(D-1)}\!S_{\diamondsuit_{1,(D-1)}}}.
\end{equation}
Recall that the action of the standard flux operator reads
 \begin{equation}
  \hat{\pi}^{IJ}({}^{(D-1)}\!S_{\diamondsuit_{(D-1)}})\cdot T_{\gamma, {}^{(D-1)}\!S_{\diamondsuit_{1,(D-1)}}}=2\mathbf{i}\hbar\kappa\beta R_{e_0}^{IJ}\cdot T_{\gamma, {}^{(D-1)}\!S_{\diamondsuit_{1,(D-1)}}}.
\end{equation}
Therefore, the actions of $\hat{\pi}^{IJ}_{\textrm{alt}}({}^{(D-1)}\!S_{\diamondsuit_{(D-1)}})$ and  $\hat{\pi}^{IJ}({}^{(D-1)}\!S_{\diamondsuit_{(D-1)}})$ on $T_{\gamma, {}^{(D-1)}\!S_{\diamondsuit_{1,(D-1)}}}$ are equivalent up to a undetermined factor in above case.
\section{General geometric operator: Second strategy}
Another way to construct general geometric operators in all dimensional LQG is to express the de-densitized dual momentum by the momentum variable $\pi^{a}_{IJ}$ as
 \begin{equation}\label{S1}
   \sqrt{q}\pi_{a}^{IJ}\hat{=}\frac{\frac{1}{(D-1)!}\epsilon_{aa_1b_1...a_nb_n}\epsilon^{IJI_1J_1...I_nJ_n} \pi^{a_1}_{I_1K_1}\pi^{b_1 K_1}_{J_1}...\pi^{a_n}_{I_nK_n}\pi^{b_n K_n}_{J_n}}{\textrm{sgn}({\det(\pi)}){|\det(\pi)|}^{\frac{D-2}{D-1}}}
   \end{equation}
   for $D=2n+1$ is odd, where
   \begin{equation}
    \det(\pi):=\frac{1}{2D!}\epsilon_{aa_1b_1...a_nb_n}\epsilon^{IJI_1J_1...I_nJ_n}\pi^{a}_{IJ} \pi^{a_1}_{I_1K_1}{\pi^{b_1 K_1}_{J_1}}...\pi^{a_n}_{I_nK_n}{\pi^{b_n K_n}_{J_n}};
   \end{equation}
  and
  \begin{equation}\label{S2}
   \sqrt{q}\pi_{a_1I_1K_1}\hat{=}\frac{\frac{2}{(D-1)!}\epsilon_{a_1b_1...a_nb_n}V^I\epsilon_{I[I_1|J_1...I_nJ_n|} {\pi^{b_1J_1}}_{K_1]}\pi^{a_2I_2K_2}{\pi^{b_2J_2}}_{K_2}...\pi^{a_nI_nK_n}{\pi^{b_nJ_n}}_{K_n}}{{\textrm{ddet}(\pi)}^{\frac{2D-3}{2D-2}}}
   \end{equation}
   for  $D=2n$ is even, where
   \begin{equation}
     V^I:=\frac{1}{D!}\epsilon_{a_1b_1...a_nb_n}\epsilon^{II_1J_1...I_nJ_n}\pi^{a_1}_{I_1K_1} {\pi^{b_1 K_1}_{J_1}}\pi^{a_2}_{I_2K_2}{\pi^{b_2 K_2}_{J_2}}...\pi^{a_n}_{I_nK_n}{\pi^{b_n K_n}_{J_n}},
   \end{equation}
   and ${\textrm{ddet}(\pi)}:=V^IV_I$. Then, we can regularize and quantize them through the flux operators, volume operator and so on, by taking account of Eqs.\eqref{qab}, (\ref{smear q1}) and (\ref{smear q2}). This strategy is similar to that used to construct the other two versions of length operator \cite{bianchi2009length}\cite{ma2010new} in the standard (1+3)-dimensional LQG. In this section, we will firstly extend the construction of the length operator in \cite{ma2010new} to all dimensional theory, and then follow a similar strategy to construct general geometric operators.
\subsection{The second length operator in all dimensional LQG}
Let us recall the classical expression \eqref{Le=} of the length $L_e$ of a curve $e$.
The length segment $L_{e^\epsilon}$ related to an arbitrary segment $e^\epsilon$ can be re-expressed by fluxes following a partition of the neighborhood of $e^\epsilon$ in $\sigma$ as follows.
Choose a set of (D-1)-faces $({}^{D-1}\!S_1,...,{}^{D-1}\!S_i,...,{}^{D-1}\!S_{D-1})$, i.e., (D-1)-hypercubes, with coordinate volume $\epsilon^{(D-1)}$ intersecting at $e_\epsilon$. The normal co-vectors $(n^1_a,...,n^i_a,...,n^{D-1}_a)$ of these (D-1)-faces are chosen to be linearly independent so that
   \begin{equation}
   \dot{e}_\epsilon^a\epsilon_{aa_1a_2...a_{D-1}}=\epsilon_{i_1...i_{D-1}}n^{i_1}_{a_1}...n^{i_{D-1}}_{a_{D-1}}
   \end{equation}
where $\epsilon_{i_1...i_{D-1}}$ is the (D-1)-dimensional Levi-Civita symbol. Taking account of the expressions \eqref{S1} and \eqref{S2} for $\sqrt{q}\pi_{aIJ}$, we can define the smeared quantity
     \begin{eqnarray}\label{smear q1}
   &&l_{e_\epsilon,IJ}\\\nonumber
   &:=&\frac{\epsilon_{i_1...i_{D-1}}\epsilon_{IJI_1J_1...I_nJ_n} \pi^{I_1K_1}({}^{D-1}\!S^{i_1}){\pi^{J_1}}_{K_1}({}^{D-1}\!S^{i_2})...\pi^{I_nK_n}({}^{D-1}\!S^{i_{D-2}}){\pi^{J_n}}_{K_n}({}^{D-1}\!S^{i_{D-1}})} {(D-1)! V_{\square_\epsilon}^{D-2}}
   \end{eqnarray}
   for $D=2n+1$ is odd, where  $V_{\square_\epsilon}=\int_{\square_\epsilon}dx^D|\det{\pi}|^{\frac{1}{D-1}}$, and $\square_\epsilon$ is the D-hypercube which contains point $v$ and has coordinate volume $\epsilon^D$. Here $\det(\pi)$ was smeared as $\det(\pi)(p)=\pi(p,\triangle_1,...,\triangle_D)$ with
   \begin{eqnarray}\label{pip}
    \pi(p,\triangle_1,...,\triangle_D) &:=&\frac{1}{\textrm{vol}(\triangle_1)...\textrm{vol}(\triangle_D)}\int_\sigma d^Dx_1...\int_\sigma d^Dx_D  \\\nonumber
      &&  \chi_{\triangle_1}(p,x_1)\chi_{\triangle_2}(2p,x_1+x_2)...\chi_{\triangle_D}(Dp,x_1+...+x_D)\\\nonumber
      && \frac{1}{2D!}\epsilon_{aa_1b_1...a_nb_n}\epsilon^{IJI_1J_1...I_nJ_n}\pi^{a}_{IJ} \pi^{a_1}_{I_1K_1}{\pi^{b_1 K_1}_{J_1}}...\pi^{a_n}_{I_nK_n}{\pi^{b_n K_n}_{J_n}}, \end{eqnarray}
      where $\chi_{\triangle}(p,x)$ denotes the characteristic function in the coordinate $x$ of a hypercube with centre $p$, which is spanned by the D right-handed vectors $\vec{\triangle}^{\tilde{i}}:=\triangle^{\tilde{i}}\vec{v}^{\tilde{i}}, \tilde{i}=1,...D$, with $\vec{v}^{\tilde{i}}$ being a normal vector in
the frame under consideration, and has coordinate volume $\text{vol}=\triangle^1...\triangle^D\det(\vec{v}^1,...,\vec{v}^D)=\epsilon^D$. Thus one has,
\begin{equation}
\chi_{\triangle}(p,x)=\prod_{\tilde{i}=1}^D\Theta\left(\frac{\triangle^{\tilde{i}}}{2}-|<\vec{v}^{\tilde{i}} ,x-p>|\right),
\end{equation}
where $<\cdot,\cdot>$ is the standard Euclidean inner product and $\Theta(y)=1$ for $y>0$ and zero otherwise.
Also, in Eq.\eqref{pip} we used the lower indices $\triangle_I=(\triangle^{1}_I,...,\triangle^{D}_I)$ to label different hypercubes, see \cite{bodendorfer2013newiii}.
Similarly, we have the smeared quantity
  \begin{eqnarray}\label{smear q2}
   &&l_{e^\epsilon,I_1K_1}\\\nonumber
   &:=&\frac{2}{(D-1)!}\epsilon_{i_1...i_{D-1}}V^I(\square_\epsilon^{D-1})\epsilon_{I[I_1|J_1...I_nJ_n|} {\pi^{J_1}}_{K_1]}({}^{D-1}\!S^{i_1})\pi^{I_2K_2}({}^{D-1}\!S^{i_2}) {\pi^{J_2}}_{K_2}({}^{D-1}\!S^{i_3})\\\nonumber
  &&\qquad\qquad\qquad\qquad\qquad\qquad\qquad\qquad...\pi^{I_nK_n}({}^{D-1}\!S^{i_{D-2}}) {\pi^{J_n}}_{K_n}({}^{D-1}\!S^{i_{D-1}}){V_{\square_\epsilon}}^{3-2D}
   \end{eqnarray}
   for  $D=2n$ is even, where $V^I(p)$ is also smeared as $V^I(\square_\epsilon^{D-1}):=[\textrm{vol}(\square_\epsilon)]^{D-2}\int_{\square_\epsilon} V^I(p)dp^D$, with $V^I(p)=V^I(p,\triangle_1,...,\triangle_D)$ and
   \begin{eqnarray}
     V^I(p,\triangle_1,...,\triangle_D) &:=&\frac{1}{\textrm{vol}(\triangle_1)...\textrm{vol}(\triangle_D)}\int_\sigma d^Dx_1...\int_\sigma d^Dx_D  \\\nonumber
      &&  \chi_{\triangle_1}(p,x_1)\chi_{\triangle_2}(2p,x_1+x_2)...\chi_{\triangle_D}(Dp,x_1+...+x_D)\\\nonumber
      && \frac{1}{D!}\epsilon_{a_1b_1...a_nb_n}\epsilon^{II_1J_1...I_nJ_n}\pi^{a_1}_{I_1K_1} {\pi^{b_1 K_1}_{J_1}}\pi^{a_2}_{I_2K_2}{\pi^{b_2 K_2}_{J_2}}...\pi^{a_n}_{I_nK_n}{\pi^{b_n K_n}_{J_n}} \end{eqnarray}
    similar to the definition of $\pi(p,\triangle_1,...,\triangle_D)$. With these smeared quantities, we can re-express the length of a curve as
 \begin{equation}
L_{e}= \lim_{\epsilon\rightarrow0}\sum_{e^\epsilon} L_{e_\epsilon} =\lim_{\epsilon\rightarrow0}\sum_{e^\epsilon}\sqrt{\frac{1}{2}l^{IJ}_{e^\epsilon}l_{IJ}^{e^\epsilon}}.
 \end{equation}
 Correspondingly, the length operator based on this expression is given by
   \begin{equation}\label{slo}
   \hat{L}_{e}=\lim_{\epsilon\rightarrow0}\sum_{e^\epsilon}\sqrt{\frac{1}{2}\hat{l}^{IJ}_{e^\epsilon}{(\hat{l}_{e^\epsilon\!,IJ})}^{ \dagger}}.
   \end{equation}
   An alternative formulation reads
  \begin{equation}\label{aslo}
   \hat{L}_{e}=\lim_{\epsilon\rightarrow0}\sum_{e^\epsilon}\frac{1}{2}\sqrt{\frac{1}{2}(\hat{l}^{IJ}_{e^\epsilon}+{(\hat{l}_{e^\epsilon\!,IJ})}^{ \dagger})(\hat{l}^{IJ}_{e^\epsilon}+{(\hat{l}_{e^\epsilon\!,IJ})}^{ \dagger})}.
   \end{equation}
     Now we need to define the operator $\hat{l}^{IJ}_{e^\epsilon}$. In all dimensional LQG \cite{bodendorfer2013newiii}, the fluxes and volume can be promoted as operators immediately. The action of a flux operator on a  cylindrical function $f_\gamma$ reads
\begin{eqnarray}
 \hat{\pi}^{IJ}({}^{D-1}\!S^{i})  \cdot f_\gamma&=& \mathbf{i}\hbar\kappa\beta \sum_{e_\imath\in E(\gamma[{}^{D-1}\!S^{i}])}\epsilon(e_\imath,{}^{D-1}\!S^{i})R_{e_\imath}^{IJ}\cdot f_\gamma, \end{eqnarray}
 where $E(\gamma[{}^{D-1}\!S^{i}])$ denotes the collection of the edges intersecting the face ${}^{D-1}\!S^{i}$, and $R_e^{IJ}$ is the right invariant vector field on $SO(D+1)\ni h_e(A)$. The action of the volume operator is given by
\begin{eqnarray}
\hat{V}_{\square_\epsilon} \cdot f_\gamma&=&(\hbar\kappa\beta)^{\frac{D}{D-1}}\sum_{v\in V(\gamma)\cap\square_\epsilon}\hat{V}_{v,\gamma}\cdot f_\gamma,
\end{eqnarray}
where
\begin{equation}
  \hat{V}_{v,\gamma} = ( \hat{V}_{v,\gamma} ^I  \hat{V}_{I\,v,\gamma} )^{\frac{1}{2D-2}},
  \end{equation}
  with
  \begin{eqnarray}
  \hat{V}_{v,\gamma}^I&=& \frac{\mathbf{i}^D}{D!}\sum_{e_1,...,e_D\in E(\gamma),e_1\cap...\cap e_D=v}s(e_1,...,e_D)\hat{q}^I_{e_1,...,e_D},\\\nonumber
   \hat{q}^I_{e_1,...,e_D}&=&  \epsilon_{II_1J_1I_2J_2...I_nJ_n}R_{e_1}^{I_1K_1}R_{e'_1K_1}^{J_1}... R_{e_n}^{I_nK_n}R_{e'_nK_n}^{J_n},
   \end{eqnarray}
   for $D$ is even, and
\begin{eqnarray}
  \hat{V}_{v,\gamma} &=& |\frac{\mathbf{i}^D}{D!}\sum_{e_1,...,e_D\in E(\gamma),e_1\cap...\cap e_D=v}s(e_1,...,e_D)\hat{q}_{e_1,...,e_D}|^{\frac{1}{D-1}},
  \end{eqnarray}
  with
  \begin{eqnarray}
   \hat{q}_{e_1,...,e_D}&=&  \frac{1}{2}\epsilon_{IJI_1J_1I_2J_2...I_nJ_n}R_e^{IJ}R_{e_1}^{I_1K_1}R_{e'_1K_1}^{J_1}... R_{e_n}^{I_nK_n}R_{e'_nK_n}^{J_n},
   \end{eqnarray}
   for $D$ is odd, where $V(\gamma)$ is the collection of vertices of the graph $\gamma$, $s(e1,...,e_D) := \text{sgn}(\det(\dot{e}_1(v),...,\dot{e}_D(v)))$, and $v$ is the intersection point of the D-tuple of edges $(e_1, e_2,...,e_D)$. It is understood that we only sum over the D-tuples of edges which are incident at a common vertex. Similarly,  we can quantize $V^I(\square_\epsilon^{D-1})$ in \eqref{smear q2} as
 \begin{eqnarray}
 && \widehat{V}^I(\square_\epsilon^{D-1})\cdot f_\gamma\\\nonumber
  &=&(\text{vol}(\square_\epsilon))^{D-2}\sum_{v\in V(\gamma)\cap\square_\epsilon}\int_{\square_\epsilon}dp \sum_{e_1,...,e_D}\frac{(\mathbf{i}\hbar\kappa\beta)^D s(e_1,...,e_D)}{D!\textrm{vol}(\triangle_1)...\textrm{vol}(\triangle_{D-1})}\chi_{\triangle_1}(p,v)...\chi_{\triangle_{D-1}}(p,v) \hat{q}^I_{e_1,...,e_D}\cdot f_\gamma \\\nonumber
  &=& (\kappa\beta\hbar)^D\sum_{v\in V(\gamma)\cap\square_\epsilon}\hat{V}_{v,\gamma}^I\cdot f_\gamma.
\end{eqnarray}
Hence the operator $\hat{l}^{IJ}_{e^\epsilon}$ is well defined by replacing the components in its classical expression with the corresponding quantum operators. Several remarks are listed below on the replacement. Firstly, the expression involves the inverse of the local volume operator $\hat{V}_{\square_\epsilon}$ which is non-invertible as it has a huge kernel. To overcome this problem, we can introduce an operator $\widehat{V^{-1}}_{\square_\epsilon}$ similar to the ``inverse'' volume operator in (1+3)-dimensional standard LQG, which is defined as the limit
      \begin{equation}
       \widehat{V^{-1}_{\square_\epsilon}}=\lim_{\varepsilon\rightarrow0} (\hat{V}^2_{\square_\epsilon}+\varepsilon^2(l_p^{(D+1)})^{2D})^{-1} \hat{V}_{\square_\epsilon},
      \end{equation}
where $l_p^{(D+1)}$ is the Plank length in (1+D)-dimensional space-time. The existence of the ``inverses'' volume operator $\widehat{V^{-1}_{\square_\epsilon}}$ indicates that the length operator will be non-vanishing only on the vertex which does not vanish the volume operator. Secondly, although the pre-quantized smeared quantities are well-defined in some limit, they are not yet background-independent because of the existence of $\epsilon_{i_1...i_{\!D-1}}\epsilon(e_{\imath_1},{}^{D-1}\!S^{i_{\!1}}) ...\\ \epsilon(e_{\imath_{\!D-1}},{}^{D-1}\!S^{i_{\!D-1}})$. The background structure can be removed by suitably ``averaging'' the regularized operator over it following a strategy similar to the treatment of the length operator in (1+3)-dimensional standard LQG \cite{ma2010new}. Then one obtains the average of $\epsilon_{i_1...i_{\!D-1}}\epsilon(e_{\imath_1},{}^{D-1}\!S^{i_{\!1}})...\epsilon(e_{\imath_{\!D-1}},{}^{D-1}\!S^{i_{\!D-1}})$ as $k_{\textrm{av}}\cdot\varsigma(e_\epsilon,e_{\imath_1},...,e_{\imath_{\!D-1}})$, wherein $k_{\textrm{av}}$ is a constant, and $\varsigma(e_\epsilon,e_{\imath_1},...,e_{\imath_{\!D-1}})$ is the orientation function which equals $+1$ (or $-1$) if the tangential directions of $e_\epsilon,e_{\imath_1},...,e_{\imath_{\!D-1}}$ are linearly independent at a vertex $v$ that dual to $\square_\epsilon$ and oriented positively (or negatively), or zero otherwise.
Thirdly, the following non-commutative relations are generally hold,
   \begin{equation}
     [\hat{l}^{IJ}_{e^\epsilon},\hat{V}_{\square_\epsilon}]\neq0,
   \end{equation}
   where $v\in e^\epsilon$ is the vertex dual to $\square_\epsilon$, and
   \begin{equation}
[\hat{L}_{e^\epsilon_\imath},\hat{L}_{e^\epsilon_\jmath}]\neq 0,
   \end{equation}
where $e^\epsilon_\imath$ and $e^\epsilon_\jmath$ intersect at a true vertex which dual to a non-vanishing volume. This result indicates that we should choose a ``nice'' extended curve to define its length operator \cite{bianchi2009length}.

Based on the above treatment the operator $\hat{l}_{e_\epsilon,IJ}$ can be given by
\begin{eqnarray}\label{lIJ1}
\hat{l}_{e_\epsilon,IJ}\cdot f_\gamma&:=&\frac{(\mathbf{i}\kappa\beta\hbar)^{D-1}}{(D-1)!}\sum_{e_{\imath_1}}...\sum_{e_{\imath_{D-1}}}k_{\textrm{av}}\cdot\varsigma(e_\epsilon,e_{\imath_1},...,e_{\imath_{\!D-1}}) \\\nonumber &&\epsilon_{IJI_1J_1...I_nJ_n}
R_{e_{\imath_1}}^{I_1K_1}{R_{e_{\imath_2}}^{J_1}}_{K_1}... R_{e_{\imath_{\!D-2}}}^{I_nK_n}{R_{e_{\imath_{\!D-1}}}^{J_n}}_{K_n} { (\widehat{V^{-1}_{\square_\epsilon}})^{D-2}}\cdot f_\gamma,
\end{eqnarray}
for $D=2n+1$ is odd, and
\begin{eqnarray}\label{lIJ2}
\hat{l}_{e_\epsilon, I_1K_1}\cdot f_\gamma&:=&
   \frac{2(\mathbf{i}\kappa\beta\hbar)^{D-1}}{(D-1)!}\sum_{e_{\imath_1}}...\sum_{e_{\imath_{D-1}}}k_{\textrm{av}}\cdot\varsigma(e_\epsilon,e_{\imath_1},...,e_{\imath_{\!D-1}}) \\\nonumber &&\hat{V}^I(\square_\epsilon^{D-1})\epsilon_{I[I_1|J_1...I_nJ_n|} {R_{e_{\imath_1}}^{J_1}}_{K_1]}R_{e_{\imath_2}}^{I_2K_2} {R_{e_{\imath_3}}^{J_2}}_{K_2}...R_{e_{\imath_{D-2}}}^{I_nK_n} {R_{e_{\imath_{D-1}}}^{J_n}}_{K_n}{\widehat{V^{-1}_{\square_\epsilon}}}^{2D-3}\cdot f_\gamma
   ,
\end{eqnarray}
for $D=2n$ is even. The final formulation of the second length operator is given by Eqs. \eqref{slo} or \eqref{aslo}.

\subsection{The second version of general $m$-area operators}
The above procedure of constructing the length operator can be extended to construct the general geometric operators measuring the $m$-area of a $m$-dimensional surface ${}^m\!S$. By the partition ${}^m\!S=\sum_{t\in \mathbb{N}, 0\leq t\leq T}{}^m\!\overline{S}^t_{\diamondsuit_{1m}}$ of an open $m$-surface ${}^m\!S$,
        the $m$-area $\text{Ar}({}^m\!S_{\diamondsuit_{1m}})$ can be re-expressed by fluxes following a partition of the neighborhood of ${}^m\!S_{\diamondsuit_{1m}}$ in $\sigma$ as follows. Suppose that the $(D-m)$ -tuple of (D-1)-surface ${}^{D-1}\!S_{i}$ ($i=1,...,D-m$) with coordinate (D-1)-area $\epsilon^{D-1}$ intersect at the $m$-dimensional region ${}^m\!S_{\diamondsuit_{1m}}$. The normal co-vectors $(n_a^1,...,n_a^i,...,n_a^{D-m})$ of ${}^m\!S_{\diamondsuit_{1m}}$ span a $(D-m)$-dimensional vector space and satisfy
\begin{equation}
\epsilon^{\imath_1...\imath_m}\dot{e}_{\imath_1}^{a_1}... \dot{e}_{\imath_m}^{a_m} =\frac{1}{(D-m)!}\epsilon^{a_1...a_ma_{m+1}...a_D}n_{a_{m+1}}^{i_1}...n_{a_{D}}^{i_{D-m}}\epsilon_{i_1...i_{D-m}}.
\end{equation}
We consider the following two cases.

\textbf{Case I: $\bar{m}:=D-m$ is even}

 Define
 \begin{eqnarray}
\bar{E}_{K_1...K_{\bar{m}}}&:=& \frac{1}{\sqrt{\bar{m}!}}\pi^{b_1}_{K_1L_1}\delta^{L_1L_2}\pi^{b_2}_{K_2L_2}... \pi^{b_{\bar{m}-1}}_{K_{\bar{m}-1}L_{\bar{m}-1}}\delta^{L_{\bar{m}-1}L_{\bar{m}}} \pi^{b_{\bar{m}}}_{K_{\bar{m}}L_{\bar{m}}}\\\nonumber
&&n_{b_1}^{i_1}...n_{b_{\bar{m}}}^{i_{\bar{m}}}\epsilon_{i_1...i_{\bar{m}}}|\det({\pi})|^{\frac{1-\bar{m}}{D-1}},
 \end{eqnarray}
 for $D$ is odd, and
 \begin{eqnarray}
\bar{E}_{K_1...K_{\bar{m}}}&:=& \frac{1}{\sqrt{\bar{m}!}}\pi^{b_1}_{K_1L_1}\delta^{L_1L_2}\pi^{b_2}_{K_2L_2}... \pi^{b_{\bar{m}-1}}_{K_{\bar{m}-1}L_{\bar{m}-1}}\delta^{L_{\bar{m}-1}L_{\bar{m}}} \pi^{b_{\bar{m}}}_{K_{\bar{m}}L_{\bar{m}}}\\\nonumber
&&n_{b_1}^{i_1}...n_{b_{\bar{m}}}^{i_{\bar{m}}}\epsilon_{i_1...i_{\bar{m}}}|\textrm{ddet}({\pi})|^{\frac{1-\bar{m}}{2D-2}},
 \end{eqnarray}
 for $D$ is even. Both of them satisfy
 \begin{equation}\label{qEE1}
\det({}^m\!q)=\bar{E}^{K_1...K_{\bar{m}}}\bar{E}_{K_1...K_{\bar{m}}},
 \end{equation}
which gives $\det{(q)}=\det{({}^m\!q)}\det{({}^{\bar{m}}\!q)}$, and
 \begin{equation}
\det{({}^{\bar{m}}\!q)^{-1}}:=\frac{1}{\bar{m}!}n_{a_1}^{i'_1}...n_{a_{\bar{m}}}^{i'_{\bar{m}}}\epsilon_{i_1...i_{\bar{m}}}q^{a_1b_1}...q^{a_{\bar{m}}b_{\bar{m}}} n_{b_1}^{i_1}...n_{b_{\bar{m}}}^{i_{\bar{m}}}\epsilon_{i_1...i_{\bar{m}}}.
 \end{equation}

\textbf{Case II: $\bar{m}:=D-m$ is odd}

Similar to last case, we can define
\begin{eqnarray}
\bar{E}_{IJK_1...K_{\bar{m}-1}}&:=& \frac{1}{\sqrt{2\bar{m}!}}\pi^{b}_{IJ}\pi^{b_1}_{K_1L_1}\delta^{L_1L_2}\pi^{b_2}_{K_2L_2}... \pi^{b_{\bar{m}-2}}_{K_{\bar{m}-2}L_{\bar{m}-2}}\delta^{L_{\bar{m}-2}L_{\bar{m}-1}} \pi^{b_{\bar{m}-1}}_{K_{\bar{m}-1}L_{\bar{m}-1}}\\\nonumber
&&n_{b}^{i}n_{b_1}^{i_1}...n_{b_{\bar{m}-1}}^{i_{\bar{m}-1}}\epsilon_{ii_1...i_{\bar{m}}}|\det({\pi})|^{\frac{1-\bar{m}}{D-1}},
 \end{eqnarray}
 for $D$ is odd, and
 \begin{eqnarray}
\bar{E}_{IJK_1...K_{\bar{m}-1}}&:=& \frac{1}{\sqrt{2\bar{m}!}}\pi^{b}_{IJ}\pi^{b_1}_{K_1L_1}\delta^{L_1L_2}\pi^{b_2}_{K_2L_2}... \pi^{b_{\bar{m}-2}}_{K_{\bar{m}-2}L_{\bar{m}-2}}\delta^{L_{\bar{m}-2}L_{\bar{m}-1}} \pi^{b_{\bar{m}-1}}_{K_{\bar{m}-1}L_{\bar{m}-1}}\\\nonumber
&&n_{b}^{i}n_{b_1}^{i_1}...n_{b_{\bar{m}-1}}^{i_{\bar{m}-1}}\epsilon_{ii_1...i_{\bar{m}}} |\textrm{ddet}({\pi})|^{\frac{1-\bar{m}}{2D-2}},
 \end{eqnarray}
 for $D$ is even. They also satisfy
 \begin{equation}\label{qEE2}
\det({}^m\!q)=\bar{E}^{IJK_1...K_{\bar{m}-1}}\bar{E}_{IJK_1...K_{\bar{m}-1}}.
 \end{equation}

 Similar to the construction of length operator,  we define
 \begin{eqnarray}\label{mathcalE1}
\bar{\mathcal{E}}^{K_1...K_{\bar{m}}}&:=& \frac{1}{\epsilon^{m}\sqrt{\bar{m}!}}\pi^{K_1L_1}({}^{D-1}\!S^{i_1})\delta_{L_1L_2}\pi^{K_2L_2}({}^{D-1}\!S^{i_2})...\\\nonumber
&&\pi^{K_{\bar{m}-1}L_{\bar{m}-1}}({}^{D-1}\!S^{i_{\bar{m}-1}})\delta_{L_{\bar{m}-1}L_{\bar{m}}} \pi^{K_{\bar{m}}L_{\bar{m}}}({}^{D-1}\!S^{i_{\bar{m}}})\epsilon_{i_1...i_{\bar{m}}}V_{\square_\epsilon}^{(1-\bar{m})},
 \end{eqnarray}
 for $\bar{m}$ is even, and
  \begin{eqnarray}\label{mathcalE2}
\bar{\mathcal{E}}^{IJK_1...K_{\bar{m}-1}}&:=& \frac{1}{\sqrt{2\bar{m}!}}\pi^{IJ}({}^{D-1}\!S^{i})\pi^{K_1L_1}({}^{D-1}\!S^{i_1})\delta_{L_1L_2}\pi^{K_2L_2}({}^{D-1}\!S^{i_2})...\\\nonumber
&&\pi^{K_{\bar{m}-2}L_{\bar{m}-2}}({}^{D-1}\!S^{i_{\bar{m}-2}})\delta_{L_{\bar{m}-2}L_{\bar{m}-1}} \pi^{K_{\bar{m}-1}L_{\bar{m}-1}}({}^{D-1}\!S^{i_{\bar{m}-1}})\epsilon_{ii_1...i_{\bar{m}-1}}V_{\square_\epsilon}^{(1-\bar{m})},
 \end{eqnarray}
 for $\bar{m}$ is odd, where $\square_\epsilon$ is a D-dimensional box with coordinate volume $\epsilon^D$ containing the tuple of ${}^{D-1}\!S^{i}$. Then, the $m$-area $\textrm{Ar}({}^m\!S)$ can be re-expressed as
 \begin{equation}\label{marea1}
{\textrm{Ar}}({}^m\!S) =\lim_{\epsilon\rightarrow0}\sum_{{}^m\!S_{\diamondsuit_{1m}}}{\textrm{Ar}}({}^m\!S_{\diamondsuit_{1m}}) =\lim_{\epsilon\rightarrow0}\sum_{{}^m\!S_{\diamondsuit_{1m}}}\sqrt{{\bar{\mathcal{E} }}^{K_1...K_{\bar{m}}}{\bar{\mathcal{E}}}_{K_1...K_{\bar{m}}}}
\end{equation}
for $\bar{m}$ is even, and
\begin{equation}\label{marea2}
\textrm{Ar}({}^m\!S) =\lim_{\epsilon\rightarrow0}\sum_{{}^m\!S_{\diamondsuit_{1m}}}{\textrm{Ar}}({}^m\!S_{\diamondsuit_{1m}}) =\lim_{\epsilon\rightarrow0}\sum_{{}^m\!S_{\diamondsuit_{1m}}}\sqrt{{ \bar{\mathcal{E}}}^{IJK_1...K_{\bar{m}-1}}{\bar{\mathcal{E}}}_{IJK_1...K_{\bar{m}-1}}}
\end{equation}
for $\bar{m}$ is odd.
Since all the components in Eqs. \eqref{mathcalE1} and \eqref{mathcalE2} have clear quantum analogues, we can obtain the general geometric operators as
  \begin{equation}\label{AmS1}
\widehat{\textrm{Ar}}({}^m\!S) =\lim_{\epsilon\rightarrow0}\sum_{{}^m\!S_{\diamondsuit_{1m}}}\widehat{\textrm{Ar}}({}^m\!S_{\diamondsuit_{1m}}) =\lim_{\epsilon\rightarrow0}\sum_{{}^m\!S_{\diamondsuit_{1m}}}\sqrt{\widehat{\bar{\mathcal{E} }}^{K_1...K_{\bar{m}}}\widehat{\bar{\mathcal{E}}}_{K_1...K_{\bar{m}}}^\dagger}
\end{equation}
for $\bar{m}$ is even, and
\begin{equation}\label{AmS2}
\widehat{\textrm{Ar}}({}^m\!S) =\lim_{\epsilon\rightarrow0}\sum_{{}^m\!S_{\diamondsuit_{1m}}}\widehat{\textrm{Ar}}({}^m\!S_{\diamondsuit_{1m}}) =\lim_{\epsilon\rightarrow0}\sum_{{}^m\!S_{\diamondsuit_{1m}}}\sqrt{\widehat{ \bar{\mathcal{E}}}^{IJK_1...K_{\bar{m}-1}}\widehat{\bar{\mathcal{E}}}^\dagger_{IJK_1...K_{\bar{m}-1}}}
\end{equation}
for $\bar{m}$ is odd. Also, an alternative formulation can be given as
 \begin{equation}\label{AmS3}
\widehat{\textrm{Ar}}({}^m\!S) =\lim_{\epsilon\rightarrow0}\sum_{{}^m\!S_{\diamondsuit_{1m}}}\widehat{\textrm{Ar}}({}^m\!S_{\diamondsuit_{1m}}) =\lim_{\epsilon\rightarrow0}\sum_{{}^m\!S_{\diamondsuit_{1m}}}\frac{1}{2}\sqrt{(\widehat{\bar{\mathcal{E}}}_\epsilon^{K_1...K_{\bar{m}}}+ {\widehat{\bar{\mathcal{E}}}}^\dagger_{K_1...K_{\bar{m}}})(\widehat{\bar{\mathcal{E}}}_\epsilon^{K_1...K_{\bar{m}}}+ {\widehat{\bar{\mathcal{E}}}}^\dagger_{K_1...K_{\bar{m}}})}
\end{equation}
for $\bar{m}$ is even, and
\begin{equation}\label{AmS4}
\widehat{\textrm{Ar}}({}^m\!S) =\lim_{\epsilon\rightarrow0}\sum_{{}^m\!S_{\diamondsuit_{1m}}}\widehat{\textrm{Ar}}({}^m\!S_{\diamondsuit_{1m}}) =\lim_{\epsilon\rightarrow0}\sum_{{}^m\!S_{\diamondsuit_{1m}}}\frac{1}{2} \sqrt{(\widehat{\bar{\mathcal{E}}}^{IJK_1...K_{\bar{m}-1}} +\widehat{\bar{\mathcal{E}}}^\dagger_{IJK_1...K_{\bar{m}-1}})(\widehat{\bar{\mathcal{E}}}^{IJK_1...K_{\bar{m}-1}} +\widehat{\bar{\mathcal{E}}}^\dagger_{IJK_1...K_{\bar{m}-1}})}
\end{equation}
for $\bar{m}$ is odd. Note that we defined
\begin{eqnarray}
\widehat{\bar{\mathcal{E}}}^{K_1...K_{\bar{m}}}&:=& \frac{1}{\sqrt{\bar{m}!}}\hat{\pi}^{K_1L_1}({}^{D-1}\!S^{i_1})\delta_{L_1L_2}\hat{\pi}^{K_2L_2}({}^{D-1}\!S^{i_2})...\\\nonumber
&&\hat{\pi}^{K_{\bar{m}-1}L_{\bar{m}-1}}({}^{D-1}\!S^{i_{\bar{m}-1}})\delta_{L_{\bar{m}-1}L_{\bar{m}}} \hat{\pi}^{K_{\bar{m}}L_{\bar{m}}}({}^{D-1}\!S^{i_{\bar{m}}})\epsilon_{i_1...i_{\bar{m}}}{\widehat{V_{\square_\epsilon}^{-1}}}^{(\bar{m}-1)}\\\nonumber
&=& \frac{(\mathbf{i}\hbar\kappa\beta)^{\bar{m}}}{\sqrt{\bar{m}!}}\sum_{e_{\imath_1},...,e_{\imath_{\!\bar{m}}}}\epsilon(e_{\imath_1},{}^{D-1}\!S^{i_{\!1}})... \epsilon(e_{\imath_{\!\bar{m}}},{}^{D-1}\!S^{i_{\!\bar{m}}}) ...\\\nonumber
&&R_{e_{\imath_1}}^{K_1L_1}\delta_{L_1L_2}R_{e_{\imath_2}}^{K_2L_2}...R_{e_{\imath_{\bar{m}-1}}}^{K_{\bar{m}-1}L_{\bar{m}-1}}\delta_{L_{\bar{m}-1}L_{\bar{m}}} R_{e_{\imath_{\bar{m}}}}^{K_{\bar{m}}L_{\bar{m}}}\epsilon_{i_1...i_{\bar{m}}}{\widehat{V_{\square_\epsilon}^{-1}}}^{(\bar{m}-1)},
 \end{eqnarray}
 for $\bar{m}$ is even, and
  \begin{eqnarray}
\widehat{\bar{\mathcal{E}}}^{IJK_1...K_{\bar{m}-1}}&:=& \frac{1}{\sqrt{2\bar{m}!}}\hat{\pi}^{IJ}({}^{D-1}\!S^{i})\hat{\pi}^{K_1L_1}({}^{D-1}\!S^{i_1})\delta_{L_1L_2}\pi^{K_2L_2}({}^{D-1}\!S^{i_2})...\\\nonumber
&&\hat{\pi}^{K_{\bar{m}-2}L_{\bar{m}-2}}({}^{D-1}\!S^{i_{\bar{m}-2}})\delta_{L_{\bar{m}-2}L_{\bar{m}-1}} \hat{\pi}^{K_{\bar{m}-1}L_{\bar{m}-1}}({}^{D-1}\!S^{i_{\bar{m}-1}})\epsilon_{ii_1...i_{\bar{m}-1}}{\widehat{V_{\square_\epsilon}^{-1}}}^{(\bar{m}-1)}\\\nonumber
&=& \frac{(\mathbf{i}\hbar\kappa\beta)^{\bar{m}}}{\sqrt{2\bar{m}!}}\sum_{e_{\imath},e_{\imath_1},...,e_{\imath_{\!\bar{m}-1}}}\epsilon(e_{\imath},{}^{D-1}\!S^{i})\epsilon(e_{\imath_1},{}^{D-1}\!S^{i_{\!1}})... \epsilon(e_{\imath_{\!\bar{m}-1}},{}^{D-1}\!S^{i_{\!\bar{m}-1}})\\\nonumber
&&R_{e_{\imath}}^{IJ}R_{e_{\imath_1}}^{K_1L_1}\delta_{L_1L_2}R_{e_{\imath_2}}^{K_2L_2}...R_{e_{\imath_{\bar{m}-2}}}^{K_{\bar{m}-2}L_{\bar{m}-2}}\delta_{L_{\bar{m}-2}L_{\bar{m}-1}} R_{e_{\imath_{\bar{m}-1}}}^{K_{\bar{m}-1}L_{\bar{m}-1}}\epsilon_{ii_1...i_{\bar{m}-1}}{\widehat{V_{\square_\epsilon}^{-1}}}^{(\bar{m}-1)},
 \end{eqnarray}
 for $\bar{m}$ is odd.
Here we can also remove the background structure by suitably
averaging the regularized operators.
The average of
$\epsilon_{i_1...i_{\!\bar{m}}}\epsilon(e_{\imath_1},{}^{D-1}\!S^{i_{\!1}})...\epsilon(e_{\imath_{\!\bar{m}}},{}^{D-1}\!S^{i_{\!\bar{m}}})$ gives ${}^{\bar{m}}\!k_{\textrm{av}}\cdot\varsigma(e_1^\epsilon,...,e_m^\epsilon,e_{\imath_1},...,e_{\imath_{\!\bar{m}}})$ for $\bar{m}$ is even, and that of
$\epsilon_{ii_1...i_{\!\bar{m}-1}}\epsilon(e_{\imath},{}^{D-1}\!S^{i})\epsilon(e_{\imath_1},{}^{D-1}\!S^{i_{\!1}})...\epsilon(e_{\imath_{\!\bar{m}-1}},{}^{D-1}\!S^{i_{\!\bar{m}-1}})$ gives ${}^{\bar{m}}\!k_{\textrm{av}}\cdot\varsigma(e_1^\epsilon,...,e_m^\epsilon,e_{\imath},e_{\imath_1},...,e_{\imath_{\!\bar{m}-1}})$  for $\bar{m}$ is odd, wherein ${}^{\bar{m}}\!k_{\textrm{av}}$ is a constant, $(e_1^\epsilon,...,e_m^\epsilon)$ is the set of edges to give ${}^m\!S_{\diamondsuit_{1m}}$, and $\varsigma(e_1^\epsilon,...,e_m^\epsilon,e_{\imath_1},...,e_{\imath_{\!\bar{m}}})$ or $\varsigma(e_1^\epsilon,...,e_m^\epsilon,e_{\imath},e_{\imath_1},...,e_{\imath_{\!\bar{m}-1}})$ is the orientation function.

We have constructed the background-independent ``elementary'' general geometric operators in all dimensional LQG. The operators \eqref{AmS1}, \eqref{AmS2}, \eqref{AmS3} and \eqref{AmS4} are symmetric. The overall undetermined factor ${}^{\bar{m}}\!k_{\textrm{av}}$ is expected to be fixed by semi-classical consistency. It should be noted that in the special case of $\bar{m}=D-1$ the general geometric operators become some length operators. However, they are not exactly the same as \eqref{slo} and \eqref{aslo}. Nevertheless, the two versions of length operators can be identified by certain operator re-ordering. Also, the standard (D-1)-area operator can be given as the special case of $\bar{m}=1$ from the general geometric operators, and the standard D-volume operator can be given as the special case of $\bar{m}=0$. Thus the construction strategy of general geometric operators is the extension of those for the standard (D-1)-area operator and D-volume operator.

It is easy to see that the elementary geometric operator $\widehat{\textrm{Ar}}({}^m\!S_{\diamondsuit_{1m}}) $ does not commute with the D-volume operator $\hat{V}_{\square_\epsilon}$ if they both contain a same vertex $v$. This implies that these elementary geometric operators are  generally non-commutative,
\begin{equation}
[\widehat{\textrm{Ar}}({}^m\!S_{\diamondsuit_{1m}}) ,\widehat{\textrm{Ar}}({}^m\!S'_{\diamondsuit_{1m}}) ]\neq0,
\end{equation}
for ${}^m\!S_{\diamondsuit_{1m}}$ and ${}^m\!S'_{\diamondsuit_{1m}}$ contain the same vertex $v$ which dual to $\square_\epsilon$. Hence we can only define the $m$-area operator of ``nice'' extended $m$-surfaces based on the ``elementary'' geometric operators as suggested in Ref.\cite{bianchi2009length}. Also, we leave the operator ordering  issue of our general geometric operators for further study \cite{Gaoping2019coherentintertwiner}.

\section{Concluding remarks}
In the previous sections, we constructed two kinds of length operators for all dimensional LQG by extending the constructions in standard (1+3)-dimensional LQG. Based on the two different strategies, we also constructed two kinds of general geometric operators to measure arbitrary $m$-areas in all dimensional LQG. In the first strategy, by Eq.\eqref{qpi1} the de-densitized dual momentum $\sqrt{q}\pi_{a}^{IJ}$ is regularized as Eq.\eqref{pi1}. Then the general geometric quantities with $\underline{\pi}(e_\epsilon)$ as building blocks can be quantized by this regularization and suitable choices of operator ordering. In the second strategy, as the de-densitized dual momentum can be expressed by the momentum $\pi^{a}_{IJ}$ and the volume element by Eqs. \eqref{S1} and \eqref{S2}, it can also be regularized as Eqs.  (\ref{smear q1}) and (\ref{smear q2}). For the general geometric quantities, the $m$-area element can be regularized by the flux of $\pi^{a}_{IJ}$ through Eqs.  (\ref{marea1}) and (\ref{marea2}). Then they can be quantized by the regularization and introducing the ``inverse'' volume operator. To get well-defined and background-independent general geometric operators, the averaging of the regularizations has to be also introduced.

Several remarks on the two kinds of general geometric operators are listed in order.
 Firstly, the first kind of general geometric operators was constructed in section 2 with the so-called (de-densitized) dual momentum, whose smeared version was expressed by the holonomy of connection. This construction would lead to some problem if the simplicity constraint was taken into account, since the action of a holonomy could change a state satisfying the constraint into non-satisfying one. To solve the problem, some projection operators should be introduced in the construction. Different from the first one, the second kind of general geometric operators constructed in section 4 would have a well behaviour even if the simplicity constraint was considered, since this kind of operators and the simplicity constraint are both totally composed of the flux operators. In this sense, the second kind of general geometric operators is expected to be a better choice than the first one in the consideration of obtaining the semi-classical spatial geometry from all dimensional LQG. Secondly, the second kind of general geometric operators contains the standard (D-1)-area operator and D-volume operator as some special cases. Hence, its construction could be regarded as a natural extension of those of standard (D-1)-area operator and D-volume operator. Different from the second one, the construction of the first kind of general geometric operators is completely different from those of standard (D-1)-area and D-volume operators. Thus it deserves checking the consistency between them in future work. Note that a similar consistency check was performed in (1+3)-dimensional standard LQG \cite{yang2016new}. Thirdly, in the construction of the first kind of general geometric operators , the choice of the operator ordering is inspired by that of alternative flux operator in (1+3)-dimensional standard LQG \cite{giesel2006consistencyi}\cite{giesel2006consistencyii}. The consistency between the alternative flux operator $\hat{\pi}^{IJ}_{\textrm{alt.}}({}^{(D-1)}\!S_{\diamondsuit_{(D-1)}})$ and the standard flux operator $\hat{\pi}^{IJ}({}^{(D-1)}\!S_{\diamondsuit_{(D-1)}})$  in (1+D)-dimensional LQG was checked in section III.

Moreover, the properties of these general geometric operators are worth further studying. Though it is hard to obtain the spectra of the general geometric operators, one may consider the semi-classical behaviour of these operators. For instance, one can study the actions of the general geometric operators on the semiclassical states that equipped with the simple coherent intertwiners \cite{Gaoping2019polytopes}. The undetermined regularization constants in these general geometric operators are also expected to be fixed in such kind of semi-classical consistency check.

\section*{Acknowledgments}
We benefited greatly from our numerous discussions with Norbert Bodendorfer, Shupeng Song and Cong Zhang. This work is supported by the National Natural Science Foundation of China (NSFC) with Grants No. 11875006 and No. 11961131013.

\bibliographystyle{unsrt}

\bibliography{geooperatorref}

\end{document}